\documentclass[12pt]{article}
\pdfoutput=1

\usepackage[a4paper,text={16.8cm,22.4cm}]{geometry}
\usepackage{amsmath,amsfonts,braket,slashed,amssymb,bm,bbm,psfrag,graphicx,color,dsfont,euscript}
\usepackage{mathtools}
\usepackage{multicol}
\usepackage{ctable}
\usepackage[small,labelfont=bf]{caption}
\usepackage{siunitx}
\usepackage{subfigure}
\usepackage{slashed}
\usepackage{authblk}
\usepackage[hidelinks]{hyperref}

\RequirePackage[sort&compress,square,comma,numbers]{natbib}
\allowdisplaybreaks
\newcommand{\F}{{\EuScript F}}
\newcommand{\J}{{\EuScript J}}

\newcommand{\G}{{\EuScript G}}
\newcommand{\X}{{\EuScript X}}
\newcommand{\vsl}{\rlap{\hspace{0.25mm}/}{v}}
\newcommand{\nsl}{\rlap{\hspace{0.25mm}/}{n}}

\newcommand{\spac}{{\hspace{0.3mm}}}

\DeclarePairedDelimiterX\MeijerM[3]{\lparen}{\rparen}%
{\begin{smallmatrix}#1 \\[1mm] #2\end{smallmatrix}\delimsize\vert\,#3}

\newcommand\MeijerG[8][]{G^{\,#2,#3}_{#4,#5}\MeijerM[#1]{#6}{#7}{#8}}
\newmuskip\pFqmuskip
\newcommand*\pFq[6][8]{  \begingroup 
  \pFqmuskip=#1mu\relax
  \mathchardef\normalcomma=\mathcode`,
  \mathcode`\,=\string"8000
  \begingroup\lccode`\~=`\,
  \lowercase{\endgroup\let~}\pFqcomma
  {}_{#2}F_{#3}{\left(\genfrac..{0pt}{}{#4}{#5};#6\right)}%
  \endgroup
}
\newcommand{\pFqcomma}{{\normalcomma}\mskip\pFqmuskip}

\begin{document}

\begin{titlepage}

\begin{flushright}
\normalsize
MITP/22-024\\ 
March 15, 2022
\end{flushright}

\vspace{1.0cm}
\begin{center}
\Large\bf
Factorization and Sudakov Resummation in Leptonic Radiative $\bm{B}\spac$ Decay -- A Reappraisal
\end{center}

\vspace{0.5cm}

\begin{center}
Anne Mareike Galda$^a$, Matthias Neubert$^{a,b}$ and Xing Wang$^a$\\
\vspace{0.7cm} 
{\sl${}^a$PRISMA$^+$ Cluster of Excellence \& Mainz Institute for Theoretical Physics\\
Johannes Gutenberg University, 55099 Mainz, Germany\\[3mm]
${}^b$Department of Physics \& LEPP, Cornell University, Ithaca, NY 14853, U.S.A.}
\end{center}
\vspace{0.8cm}
\begin{abstract}
The $B$-meson light-cone distribution amplitude is an important non-perturbative quantity arising in the factorization of the amplitudes for many exclusive decays of $B$ mesons, such as $B^-\to\gamma\,\ell^-\spac\bar\nu$. We reconsider the renormalization-group (RG) equation satisfied by this function and present its solution at next-to-leading order (NLO) in RG-improved perturbation theory in Laplace space and, for the first time, in momentum space and the so-called diagonal (or dual) space. Since the information needed to describe the $B$ decay processes at leading order in $\Lambda_{\rm QCD}/m_b$ is most directly contained in the distribution amplitude in Laplace space evaluated near the origin, we propose an unbiased parameterization of this object in terms of a small set of uncorrelated hadronic parameters. Using recent results on the three-loop anomalous dimension for heavy-light current operators, we derive an expression for the convolution integral appearing in the $B^-\to\gamma\,\ell^-\spac\bar\nu$ factorization formula that is explicitly scale independent, and we evaluate this formula at (approximate) NNLO.
\end{abstract}
\end{titlepage}

\section{Introduction}

Studies of rare exclusive $B$-meson decays are an essential tool to test the flavor sector of the Standard Model. In order to match the accuracy of the experiments, there is a growing need for precise theoretical predictions of the relevant decay rates. The QCD factorization approach allows one to perform model-independent calculations of exclusive (quasi) two-body decay amplitudes of $B$ mesons in the heavy-quark limit \cite{Beneke:1999br,Beneke:2000ry,Beneke:2001ev,Beneke:2003zv}. The non-perturbative input required for such calculations are meson decay constants, $B$-meson transition form factors, and light-cone distribution amplitudes (LCDAs) of the $B$ meson and the light final-state mesons in the decay process. While decay constants and transition form factors can, at least in principle, be extracted using experimental data, the LCDAs describe the inner structure of the mesons involved and can at best be constrained using experimental information. Theoretical information on LCDAs can be obtained using (light-cone) QCD sum rules \cite{Grozin:1996pq,Ball:2003fq,Braun:2003wx}. Alternatively, significant progress has recently been made based on lattice QCD and the formalism of parton pseudo-distribution to study LCDAs \cite{Wang:2019msf,Zhao:2020bsx}. While LCDAs are genuinely non-perturbative quantities, their scale dependence is calculable in perturbation theory, just like in the case of parton distribution functions. 

\begin{figure}
\centering
\includegraphics[width=4.5cm]{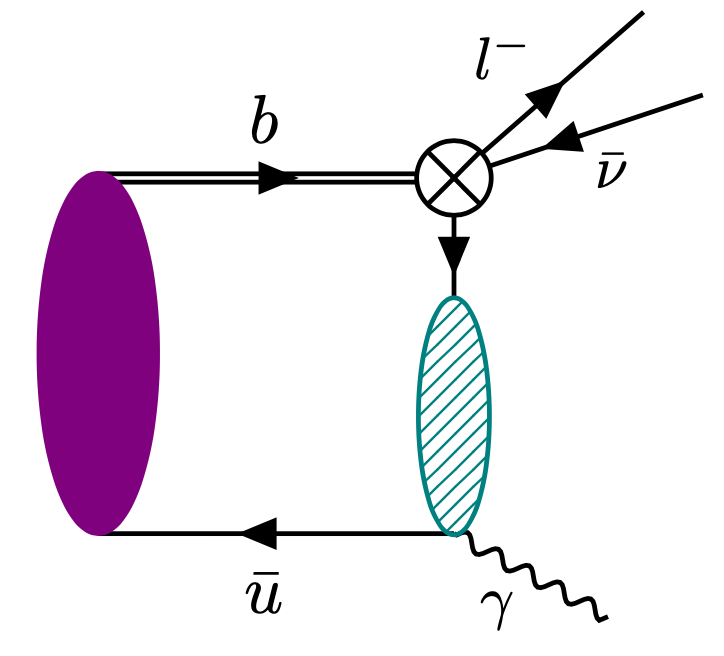}
\caption{\label{fig:factorization} 
Factorization of the $B^-\to\gamma\,\ell^-\spac\bar\nu$ decay amplitude in the region where $E_\gamma\lesssim m_B/2$. The crossed dot represents the hard function $H$, while the hatched green ellipse denotes the radiative jet function $J$ with an external photon. The $B$-meson LCDA is represented by the violet ellipse. The double line denotes a soft heavy-quark field in HQET.}
\end{figure}

The radiative leptonic decay $B^-\to\gamma\,\ell^-\spac\bar\nu$ is of particular interest. On the one hand, this process is a background to the semileptonic decay $B^-\to\ell^-\,\bar\nu$, which can be used to determine the CKM matrix element $|V_{ub}|$ \cite{Becirevic:2009aq}. A precise control of the background is a prerequisite to a reliable extraction of $|V_{ub}|$. On the other hand, because of its simplicity and the fact that no hadron is contained in the final state, in the limit of large photon energy the decay $B^-\to\gamma\,\ell^-\spac\bar\nu$ can be used to extract valuable information about moments of the leading-order LCDA $\phi_+^B(\omega,\mu)$ of the $B$ meson \cite{Lunghi:2002ju,Bosch:2003fc,Beneke:2011nf}. When the energy of the photon is large, close to its kinematic endpoint near $m_B/2$, the highly energetic photon probes the light-cone structure of the $B$ meson. At leading order (LO) in $\Lambda_{\rm QCD}/m_b$, the corresponding QCD factorization theorem reads 
\begin{equation}\label{fact}
   {\cal M}(B^-\to\gamma\,\ell^-\spac\bar\nu) 
   \propto m_B\spac f_B\,H(m_b,E_\gamma,\mu)\!\int_0^\infty\!\frac{d\omega}{\omega}\,
    J(-2E_\gamma\spac\omega,\mu)\,\phi_+^B(\omega,\mu) 
    + \mathcal{O}\bigg( \frac{\Lambda_{\text{QCD}}}{m_b} \bigg) \,,
\end{equation}
where $E_\gamma\lesssim m_B/2$ denotes the photon energy as measured in the rest frame of the $B$ meson. The hard function $H$ and the radiative jet function $J$ are calculable in perturbation theory. In particular, the jet function depends only logarithmically on the convolution variable $\omega$. This makes the decay process a clean probe of the logarithmic moments of the $B$-meson LCDA, defined in relation (\ref{eq:moments}) below. The factorization formula is illustrated in Figure~\ref{fig:factorization}. The LCDA appears due to the interactions of the high-energy (collinear) photon with the soft spectator quark inside the $B$ meson.

In contrast to the LCDAs of light mesons, not much is known on general grounds about the properties of the $B$-meson LCDA. In particular, the function $\phi_+^B(\omega,\mu)$ does not approach a simple asymptotic form in the formal limit $\mu\to\infty$. It has been shown, however, that for sufficiently large values of $\mu$ the LCDA scales like $\omega$ for $\omega\to 0$ and falls off slower than $1/\omega$ for $\omega\to\infty$ \cite{Lange:2003ff}. Several models for $\phi_+^B(\omega,\mu)$ have been proposed in the literature, which are either based on the above-mentioned QCD sum-rule estimates \cite{Grozin:1996pq,Ball:2003fq,Braun:2003wx} or invoke {\em ad hoc\/} modeling \cite{Lee:2005gza,Bell:2013tfa,Feldmann:2014ika,Beneke:2018wjp}. Most of these models rest on some unjustified assumptions, which imply important biases and lead to uncontrolled systematic uncertainties: 
\begin{enumerate}
\item 
The LCDA is often assumed to be positive definite, even though it is an {\em amplitude\/} that does not admit a probabilistic interpretation. In fact, it has been argued that $\phi_+^B(\omega,\mu)$ must change sign for some value of $\omega\gg\Lambda_{\rm QCD}$ \cite{Braun:2003wx,Lee:2005gza}. 
\item
Many models suppose that at a low renormalization scale $\mu_s$ the LCDA exhibits an exponential fall-off for large $\omega\gg\Lambda_{\rm QCD}$, even though this is in conflict with RG evolution. At best, this assumption could be true at one particular value of $\mu_s$, but RG evolution to a scale $\mu>\mu_s$ inevitably leads to a fall-off slower than $1/\omega$ \cite{Lange:2003ff}. 
\item
It is sometimes assumed that the integral over $\phi_+^B(\omega,\mu_s)$ is normalized to unity, even though this integral is, in fact, known to be divergent \cite{Grozin:1996pq}.
\end{enumerate}
In \cite{Galda:2020epp} and the present work, we argue that the information that can be probed in hard exclusive processes such as $B^-\to\gamma\,\ell^-\spac\bar\nu$ is entirely and most directly described by the {\em Laplace transform\/} of the LCDA evaluated near the origin. Describing this function by means of a simple Taylor series, we thus obtain a model-independent parameterization of the hadronic effects accessible in such processes, thereby avoiding the hidden biases introduced when a specific model for the momentum-space (or dual-space) LCDA is invoked.

In position space, the LCDA is defined as \cite{Grozin:1996pq}
\begin{equation}
   \langle 0|\,\bar q(z)\spac S_n(z,0)\,\nsl\,\Gamma\,h_v(0)|\bar B(v)\rangle 
   = -i\,\frac{F(\mu)}{2}\,\widetilde\phi_+^B(\tau,\mu)\,
    \mathrm{Tr}\!\left( \nsl\,\Gamma\,\frac{1+\vsl}{2}\,\gamma_5 \right) ,
\end{equation}
where $v$ denotes the 4-velocity of the $B$ meson, $h_v$ is the heavy-quark spinor field in heavy-quark effective theory (HQET) \cite{Georgi:1990um,Neubert:1993mb}, $n$ is a light-like reference vector in the direction of the photon (with $v\cdot n=1$), $\Gamma$ is a generic Dirac matrix, and $\tau=v\cdot z-i0$. The quantity $S_n(z,0)$ denotes a Wilson line connecting the points 0 and $z$ on a straight light-like segment. Finally, $F(\mu)$ is the $B$-meson decay constant in HQET, which is related to the decay constant $f_B$ in full QCD via \cite{Neubert:1992fk}
\begin{equation}\label{KFdef}
   \sqrt{m_B}\,f_B = K_F(m_b,\mu)\,F(\mu) + \mathcal{O}\bigg( \frac{\Lambda_{\text{QCD}}}{m_b} \bigg) \,, 
\end{equation}
with a matching coefficient $K_F$ that is known at two-loop order \cite{Broadhurst:1994se,Grozin:1998kf}. The $B$-meson LCDA in momentum space, which enters in the factorization formula \eqref{fact}, is obtained via Fourier transformation, such that \cite{Grozin:1996pq,Lange:2003ff}
\begin{equation}
   \phi_+^B(\omega,\mu) = \frac{1}{2\pi} \int\!d\tau\,e^{i\omega\tau}\,\widetilde\phi_+^B(\tau,\mu) \,.
\end{equation}

The one-loop renormalization-group (RG) equation for the leading-order $B$-meson LCDA and its analytic solution in momentum-space was derived in \cite{Lange:2003ff}. The two-loop contribution to the evolution kernel was obtained much later, first using conformal symmetry in the so-called ``dual'' space \cite{Braun:2019wyx}, in which the one-loop kernel is diagonalized using a suitable integral transformation \cite{Bell:2013tfa,Braun:2014owa}, and later in momentum space \cite{Liu:2020ydl}. In the present work we will employ techniques developed in the context of Higgs physics \cite{Liu:2020ydl,Liu:2020eqe} to construct an analytic solution of the two-loop RG equation in momentum space. In a recent letter, two of us have shown that both the evolution equation and its solution take on a much simpler form in Laplace space \cite{Galda:2020epp}. We briefly recapitulate the main features of this solution and then use it to obtain the $B$-meson LCDA in the ``diagonal'' space \cite{Liu:2020eqe}, which generalizes the concept of the dual space to higher orders of perturbation theory. In the diagonal space the evolution is local in the momentum variable $\omega$ to all orders of perturbation theory, while in the dual space it is local only at one-loop order.

The solutions of the RG evolution equation for the LCDA play an important role in the numerical evaluation of the factorization formula \eqref{fact}. The reason is that there is no common choice of the factorization scale $\mu$, for which the three functions $H$, $J$ and $\phi_+^B$ are free of large logarithmic corrections. These large logarithms can be resummed to all orders in perturbation theory by solving the evolution equations for the hard and jet functions and for the LCDA. It has been emphasized in \cite{Galda:2020epp} that a particularly elegant way of performing this resummation is obtained in Laplace space, where it is possible to construct a resummed formula that is {\em explicitly\/} independent of the choice of $\mu$. In the last part of this paper, we give a detailed discussion of this solution, in which all relevant technical details are presented. At fixed order in perturbation theory, the $B^-\to\gamma\,\ell^-\spac\bar\nu$ decay amplitude in the heavy-quark limit can be parameterized in terms of the first inverse moment $\lambda_B^{-1}$ of the LCDA and related logarithmic moments $\sigma_n^B$, which are defined as \cite{Beneke:2011nf}
\begin{equation}\label{eq:moments}
\begin{aligned}
   \frac{1}{\lambda_B(\mu)} &= \int_0^\infty\!\frac{d\omega}{\omega}\,\phi_+^B(\omega,\mu) \,, \\
   \sigma_n^B(\mu) &= \lambda_B(\mu) \int_0^\infty\!\frac{d\omega}{\omega}\,
    \ln^n\!\Big(\frac{\bar\omega}{\omega}\Big)\,\phi_+^B(\omega,\mu) \,; \quad n\ge 1 \,.
\end{aligned}
\end{equation}
Here $\bar\omega$ is an auxiliary reference scale, which can be chosen at will. A convenient choice is to adjust this parameter in such a way that, at the scale where the LCDA is given, its first moment $\sigma_1^B$ vanishes.\footnote{Alternatively, one could fix $\bar\omega$ to some reference scale, such as $\bar\omega=\mu_s$ \cite{Beneke:2011nf} or $\bar\omega=e^{-\gamma_E}\lambda_B(\mu_s)$ \cite{Beneke:2018wjp}, and keep $\sigma_1(\mu_s)$ as an independent parameter}.
The $B^-\to\gamma\,\ell^-\spac\bar\nu$ decay amplitude is particularly sensitive to $\lambda_B$ \cite{Lunghi:2002ju,Bosch:2003fc,Beneke:2011nf}, and an important goal is to derive information on this parameter (and some of the logarithmic moments) from future data obtained with the Belle~II experiment. An important outcome of this paper is the derivation of a coupled set of RG evolution equations for the hadronic parameters $\lambda_B$ and $\sigma_n^B$. The exact solution to these equations is derived in terms of the Laplace-space LCDA. We also argue that for the relevant region in Laplace space (close to the origin in the Laplace variable $\eta$), a {\em model-independent\/} parameterization of this function can be obtained in terms of the parameters $\lambda_B$ and $\sigma_n^B$ defined at a low scale $\mu_s$.

Our analysis in this work is limited to the leading-power contributions to the $B^-\to\gamma\,\ell^-\spac\bar\nu$ decay amplitude. There exist several sources of power-suppressed contributions, which have been studied in \cite{Beneke:2018wjp,Wang:2018wfj,Shen:2020hsp}.

\section{Two-loop RG evolution equation}

The general form of the RG evolution equation capturing the scale dependence of the $B$-meson LCDA is \cite{Lange:2003ff}
\begin{equation}\label{RGEphiB}
   \frac{d}{d\ln{\mu}}\,\phi_+^B(\omega,\mu) 
   = - \int_0^\infty\!d\omega'\,\gamma_+(\omega,\omega';\mu)\,\phi_+^B(\omega',\mu) \,,
\end{equation}
with the anomalous dimension 
\begin{equation}\label{anomalous}
   \gamma_+(\omega,\omega';\mu) 
   = \left[ \Gamma_{\mathrm{cusp}}(\alpha_s)\,\ln\frac{\mu}{\omega} + \gamma(\alpha_s) \right] \delta(\omega-\omega') 
    - \Gamma_{\mathrm{cusp}}(\alpha_s)\,\omega\,\Gamma(\omega,\omega') 
    - \hat\gamma(\omega,\omega';\alpha_s) \,.
\end{equation}
The first term is local in the variables $\omega$ and $\omega'$. It contains the so-called ``cusp logarithm'', whose coefficient is given by the light-like cusp anomalous dimension in the fundamental representation of $SU(N_c)$, given by $\Gamma_{\mathrm{cusp}}(\alpha_s)=\frac{C_F\spac\alpha_s}{\pi}+{\cal O}(\alpha_s^2)$, and known to four-loop order in perturbation theory \cite{Henn:2019swt}. The second term in \eqref{anomalous}, whose coefficient is also given by the cusp anomalous dimension, is proportional to the symmetric Lange--Neubert kernel
\begin{equation}
   \Gamma(\omega,\omega') 
   = \left[ \frac{\theta(\omega-\omega')}{\omega(\omega-\omega')} 
    + \frac{\theta(\omega'-\omega)}{\omega'(\omega'-\omega)} \right]_+ .
\end{equation}
It is defined such that, if integrated with a function $f(\omega')$, one needs to substitute $f(\omega')\to f(\omega')-f(\omega)$ under the integral. The third term, which is also non-local in $\omega$ and $\omega'$, is absent at one-loop order. Its two-loop expression was derived in \cite{Braun:2019wyx,Liu:2020ydl} and reads
\begin{equation}
   \hat\gamma(\omega,\omega';\alpha_s) 
   = C_F \left( \frac{\alpha_s}{2\pi} \right)^2 \frac{\omega\,\theta(\omega'-\omega)}{\omega'\,(\omega'-\omega)}\,
    h\Big(\frac{\omega}{\omega'}\Big) + {\cal O}(\alpha_s^3) \,,
\end{equation}
with 
\begin{equation}\label{hfundef}
   h(x) = \ln x \left[ \beta_0 + 2\spac C_F \left( \ln x - \frac{1+x}{x}\,\ln(1-x) - \frac32 \right) \right] .
\end{equation}

\begin{figure}
\centering
\includegraphics[width=8.1cm]{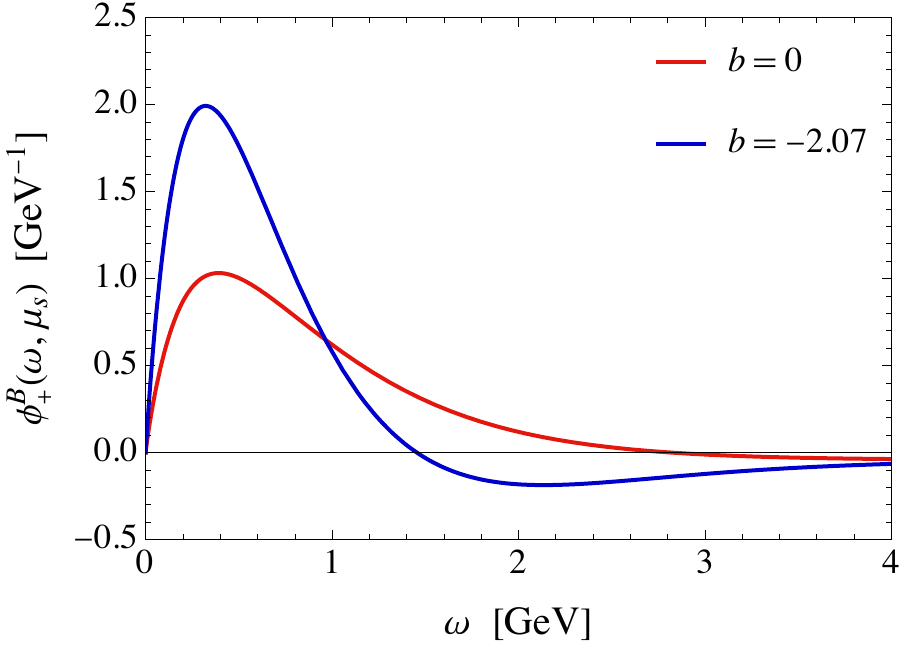}\quad
\includegraphics[width=8.1cm]{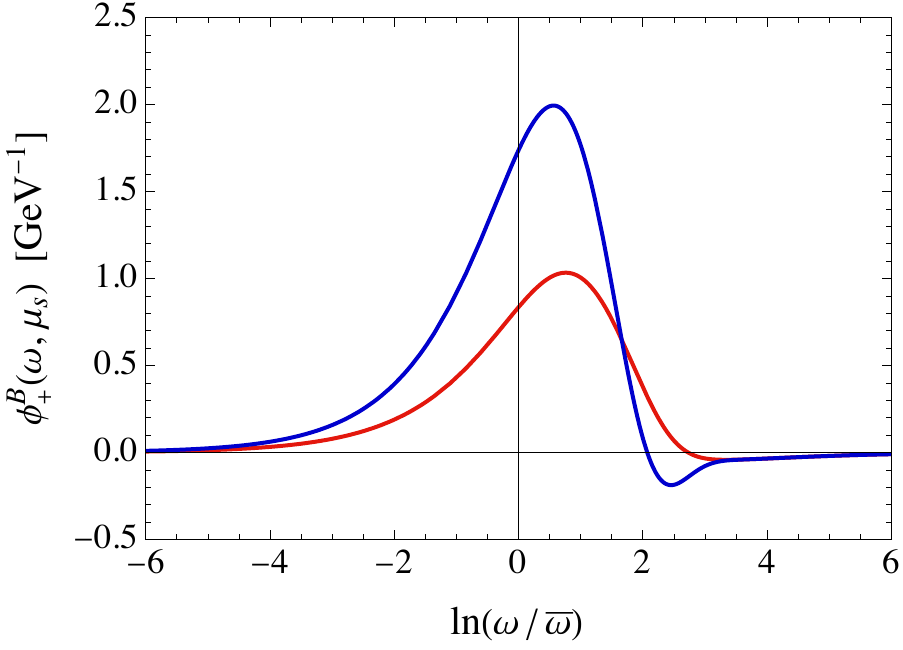}
\caption{\label{fig:LCDAmodels} 
Left: The two model functions for the LCDA $\phi_+^B(\omega,\mu_s)$ at the low scale $\mu_s=1$\,GeV, obtained with $b=0$, $\bar\omega=183$\,MeV (red curve) and $b=-2.07$, $\bar\omega=141$\,MeV (blue curve). In both cases we set $\omega_0=482$\,MeV. The parameters are chosen such that $\lambda_B(\mu_s)=350$\,MeV and 200\,MeV, respectively, whereas $\sigma_1^B(\mu_s)=0$ in both cases. Right: The two model LCDAs as functions of the variable $\ln(\omega/\bar\omega)$. The area under the curves corresponds to the parameter $1/\lambda_B(\mu_s)$, while in each case the parameter $\bar\omega$ is chosen such that the functions are centered at~0.}
\end{figure}

In order to illustrate our results in the following sections, we will employ a simple two-parameter model for the LCDA $\phi_+^B(\omega,\mu_s)$ at the low scale $\mu_s=1$\,GeV, which satisfies (most of) the known properties of the this function, such as its radiative tail for large values of $\omega$ \cite{Braun:2003wx,Lee:2005gza}. The model function reads \cite{Galda:2020epp}
\begin{equation}\label{eq:modelFunction}
   \phi_+^B(\omega,\mu_s) 
   = \left( 1 - b + \frac{b\spac\omega}{2\spac\omega_0} \right) \frac{\omega}{\omega_0^2}\,e^{-\omega/\omega_0}
    + \frac{4\alpha_s(\mu_s)}{3\pi}\,\frac{\omega}{\omega^2+\omega_0^2} 
    \left( \frac12 - \ln\frac{\omega}{\mu_s} \right) .
\end{equation}
It contains two free parameters, $b$ and $\omega_0$, which can be varied to obtain different shapes of the LCDA. The asymptotic behavior is $\phi_+^B(\omega,\mu_s)\propto\omega$ for $\omega\to 0$ and $\phi_+^B(\omega,\mu_s)\propto 1/\omega$ for $\omega\to\infty$. As mentioned earlier, we find it convenient to adjust the auxiliary scale parameter $\bar\omega$ in \eqref{eq:moments} such that $\sigma_1^B(\mu_s)=0$. In essence, we thus trade the hadronic parameter $\sigma_1^B(\mu_s)$ for a new parameter $\bar\omega\sim\Lambda_{\rm QCD}$. For the model function, setting the first moment to zero yields
\begin{equation}
   \bar\omega = \omega_0\,\exp\left[ 
    - \frac{6\gamma_E-3b(1+\gamma_E)+\pi^2\spac\alpha_s(\mu_s)}%
           {6-3b+2\alpha_s(\mu_s) \left( 1 - 2\ln\frac{\omega_0}{\mu_s} \right)} \right] .
\end{equation}
According to \eqref{eq:moments} this defines $\bar\omega$ such that the average value of the distribution amplitude $\phi_+^B(\omega,\mu_s)$ in the variable $\ln(\omega/\bar\omega)$ vanishes. With this choice it is likely that the higher moments do not take unnaturally large values either. We emphasize that the model function is used for illustrative purposes only and no claim is made that it provides a valid representation of the true LCDA. For the purposes of illustration, we keep $\omega_0=482$\,MeV fixed and consider the two choices $b=0$ and $b=-2.07$ at the reference scale $\mu_s=1$\,GeV, for which the first integral in \eqref{eq:moments} yields $\lambda_B(\mu_s)\simeq 350$\,MeV and 200\,MeV, respectively. While a value around 350\,MeV is often considered as a default choice for $\lambda_B$, in phenomenological applications of the QCD factorization approach to non-leptonic $B$ decays one typically prefers a lower value around 200\,MeV (see e.g.\ scenarios S2 and S4 in \cite{Beneke:2003zv}). The two model functions are depicted in Figure~\ref{fig:LCDAmodels} vs.\ $\omega$ (left) and $\ln(\omega/\bar\omega)$ (right). The numerical values of the first few moments of these functions are given in Table~\ref{tab:1}. Note that, due to the negative tail of the LCDA at large values of $\omega$, the fourth moment $\sigma_4^B(\mu_s)$ is negative for the two models, which would be impossible if the LCDA was a positive definite quantity.

\begin{table}
\centering
\begin{tabular}{|c|c||c|c|c|c|c|}
\hline
$b$ & $\lambda_B(\mu_s)$ & $\bar\omega$ & $\sigma_1^B(\mu_s)$ & $\sigma_2^B(\mu_s)$ & $\sigma_3^B(\mu_s)$
 & $\sigma_4^B(\mu_s)$ \\
\hline
 0 & 183\,MeV & 350\,MeV & 0 & 1.17 & 6.41 & $-6.88$ \\
\hline
 $-2.07$ & 141\,MeV & 200\,MeV & 0 & 1.04 & 5.32 & $-3.90$ \\
\hline
\end{tabular}
\caption{\label{tab:1} 
Model parameters (left) and logarithmic moments (right) of the model functions.}
\end{table}

\subsection{Solution in Laplace space}

Solving the integro-differential equation \eqref{RGEphiB} is not an easy task. An elegant all-order solution can be obtained in Laplace space. We define
\begin{equation}\label{eq:Laplacedef}
   \tilde\phi_+^B(\eta,\mu) 
   = \int_0^\infty\!\frac{d\omega}{\omega} \left( \frac{\omega}{\bar\omega} \right)^{-\eta} \phi_+^B(\omega,\mu) \,,
\end{equation}
where $\bar\omega$ serves as a fixed reference scale. It has been shown in \cite{Galda:2020epp} (see also \cite{Liu:2020eqe} for an analogous equation for the soft-quark soft function in Higgs physics) that the RG evolution equation satisfied by the Laplace-space LCDA reads
\begin{equation}\label{wonderful}
    \left( \frac{d}{d\ln\mu} + \Gamma_{\mathrm{cusp}}(\alpha_s)\,\frac{\partial}{\partial\eta} \right) 
     \tilde\phi_+^B(\eta,\mu) 
    = \left[ \Gamma_{\mathrm{cusp}}(\alpha_s) \left( \ln\frac{\bar\omega}{\mu} + \F(\eta) \right)
     - \gamma(\alpha_s) + \G(\eta,\alpha_s) \right] \tilde\phi_+^B(\eta,\mu) \,,
\end{equation}
where $\alpha_s\equiv\alpha_s(\mu)$, and we have defined 
\begin{equation}\label{eq:eigenfunctions}
\begin{aligned}
   \F(\eta) &= \int_0^\infty\!d\omega'\,\Big( \frac{\omega'}{\omega} \Big)^\eta\,\omega\,\Gamma(\omega,\omega')
    = - \big[ H(\eta) + H(-\eta) \big] \,, \\
   \G(\eta,\alpha_s)
   &= \int_0^\infty\!d\omega'\,\Big( \frac{\omega'}{\omega} \Big)^\eta\,\hat\gamma(\omega,\omega';\alpha_s) \\
   &= C_F \left( \frac{\alpha_s}{2\pi} \right)^2 \frac{d}{d\eta}\,\Big[ 
    2 C_F\spac H(-\eta)\,H(-\eta-1) - (3 C_F - \beta_0)\,H(-\eta) \Big] + {\cal O}(\alpha_s^3) \,, 
\end{aligned}
\end{equation}
where $H(\eta)=\psi(1+\eta)+\gamma_E$ is the harmonic-number function. The general solution to the evolution equation \eqref{wonderful} can be obtained by noting that any function of $\eta+a_\Gamma(\mu_s,\mu)$, where 
\begin{equation}\label{eq:aGamma}
   a_\Gamma(\mu_s,\mu) = - \int\limits_{\alpha_s(\mu_s)}^{\alpha_s(\mu)}\!d\alpha\,
    \frac{\Gamma_{\rm cusp}(\alpha)}{\beta(\alpha)} \,,
\end{equation}
provides a solution to the homogeneous equation, where the right-hand side is set to zero. One then finds that the general solution to the inhomogeneous equation is \cite{Galda:2020epp}
\begin{equation}\label{eq:LaplaceSolu}
\begin{aligned}
   \tilde\phi_+^B(\eta,\mu)
   &= \tilde\phi_+^B\big(\eta+a_\Gamma(\mu_s,\mu),\mu_s\big)\,N(\bar\omega;\mu_s,\mu)\, 
    \frac{\Gamma\big(1+\eta+a_\Gamma(\mu_s,\mu)\big)\,\Gamma(1-\eta)}%
         {\Gamma\big(1-\eta-a_\Gamma(\mu_s,\mu)\big)\,\Gamma(1+\eta)}\,
    e^{2\gamma_E\spac a_\Gamma(\mu_s,\mu)} \\
   &\quad \times \exp\!\left[\, \int\limits_{\alpha_s(\mu_s)}^{\alpha_s(\mu)}\!\frac{d\alpha}{\beta(\alpha)}\,
    \G\big(\eta+a_\Gamma(\mu_\alpha,\mu),\alpha\big) \right] ,
\end{aligned}
\end{equation}
with $\mu_\alpha$ defined such that $\alpha_s(\mu_\alpha)=\alpha$, and
\begin{equation}\label{eq:norm}
   N(\bar\omega;\mu_s,\mu) 
   = \left( \frac{\bar\omega}{\mu_s} \right)^{-a_\Gamma(\mu_s,\mu)} 
    \exp\Big[ S_\Gamma(\mu_s,\mu) + a_\gamma(\mu_s,\mu) \Big] \,.
\end{equation}
The function $a_\gamma(\mu_s,\mu)$ is defined in analogy with \eqref{eq:aGamma} but with $\Gamma_{\rm cusp}$ replaced by $\gamma$, and the Sudakov exponent is given by
\begin{equation}\label{eq:RGFunctions}
   S_\Gamma(\mu_s,\mu)
   = - \int\limits_{\alpha_s(\mu_s)}^{\alpha_s(\mu)}\!d\alpha\,
    \frac{\Gamma_{\mathrm{cusp}}(\alpha)}{\beta(\alpha)} 
    \int\limits_{\alpha_s(\mu_s)}^{\alpha} \frac{d\alpha'}{\beta(\alpha')} \,.
\end{equation}
Note that under a scale transformation the argument of the Laplace-space LCDA is shifted from $\eta$ to $\eta+a_\Gamma(\mu_s,\mu)$. 

The positions of the nearest singularities at positive and negative values of $\eta$ determine the asymptotic behavior of the momentum-space LCDA for small and large values of $\omega$ \cite{Lange:2003ff}. At the low scale $\mu_s$ we denote these values by $\eta_+$ and $-\eta_-$, respectively. The corresponding behavior of the momentum-space LCDA is $\phi_+^B(\omega,\mu_s)\propto\omega^{\eta_+}$ for $\omega\to 0$ and $\phi_+^B(\omega,\mu_s)\propto\omega^{-\eta_-}$ for $\omega\to\infty$. When the LCDA is evolved to a higher scale $\mu>\mu_s$, the positions of these singularities shift to $\eta_+ +|a_\Gamma(\mu_s,\mu)|$ and $-\eta_- +|a_\Gamma(\mu_s,\mu)|$, taking into account that $a_\Gamma(\mu_s,\mu)<0$ for $\mu>\mu_s$. Additional singularities are generated by the $\Gamma$-functions in the numerator of \eqref{eq:LaplaceSolu} and are located at $\eta=n$ and $\eta=-n+|a_\Gamma(\mu_s,\mu)|$ for all integers $n\in\mathbb{N}$. For sufficiently large values of $\mu$, the nearest positive singularity is the one at $\eta=1$, corresponding to a linear behavior $\phi_+^B(\omega,\mu)\sim\omega$ near the origin. The nearest negative singularity is located at $\eta=-\min(1,\eta_-)+|a_\Gamma(\mu_s,\mu)|$, implying that $\phi_+^B(\omega,\mu)$ falls off slower than $1/\omega$ at large~$\omega$.

At leading-order in RG-improved perturbation theory, the term shown in the second line of \eqref{eq:LaplaceSolu} can be replaced by~1. The function $\G$ starts at two-loop order, and using its definition shown in the second line of \eqref{eq:eigenfunctions} one obtains \cite{Liu:2020eqe,Galda:2020epp}
\begin{equation}\label{exponential}
   \int\limits_{\alpha_s(\mu_s)}^{\alpha_s(\mu)}\!\frac{d\alpha}{\beta(\alpha)}\,
    \G\big(\eta+a_\Gamma(\mu_\alpha,\mu),\alpha\big) 
   = \frac{C_F\spac\alpha_s(\mu)}{2\pi} \int_0^1\!\frac{dx}{1-x}\,\frac{h(x)}{\beta_0}\,x^{-\eta}\,
    \frac{r^{1+\frac{2C_F}{\beta_0} \ln x}-1}{1+\frac{2C_F}{\beta_0} \ln x} + {\cal O}(\alpha_s^2) \,,
\end{equation}
with $r=\alpha_s(\mu_s)/\alpha_s(\mu)$. The correction term enters first at next-to-leading order (NLO) in RG-improved perturbation theory. It generates additional singularities located at $\eta=n$ and $\eta=n+|a_\Gamma(\mu_s,\mu)|$
with $n\in\mathbb{N}$. Note that the nearest singularity for positive $\eta$ (for sufficiently large values of $\mu$) remains localized at $\eta=1$.

\begin{figure}
\centering
\includegraphics[width=8.1cm]{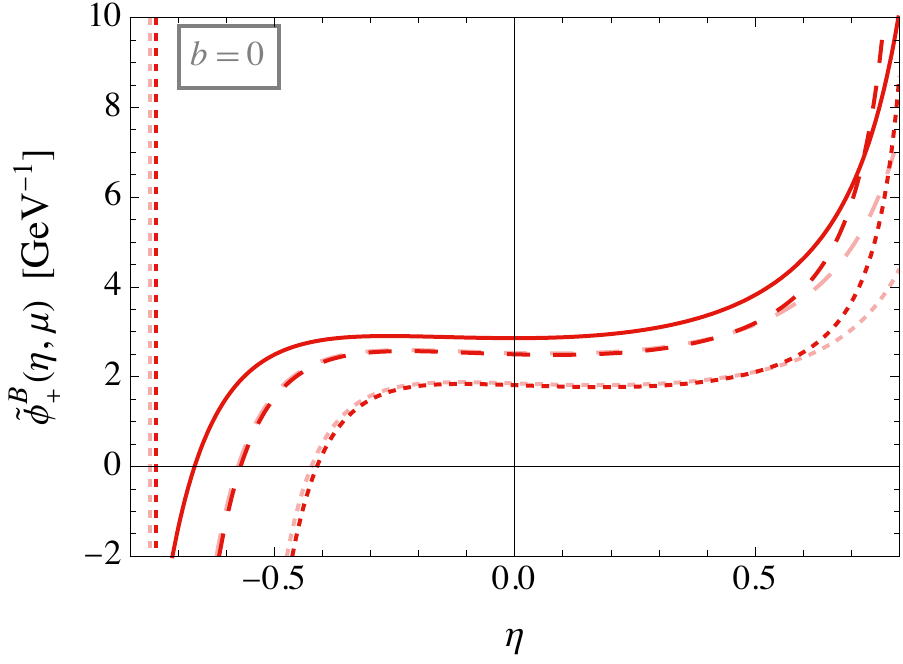}\quad
\includegraphics[width=8.1cm]{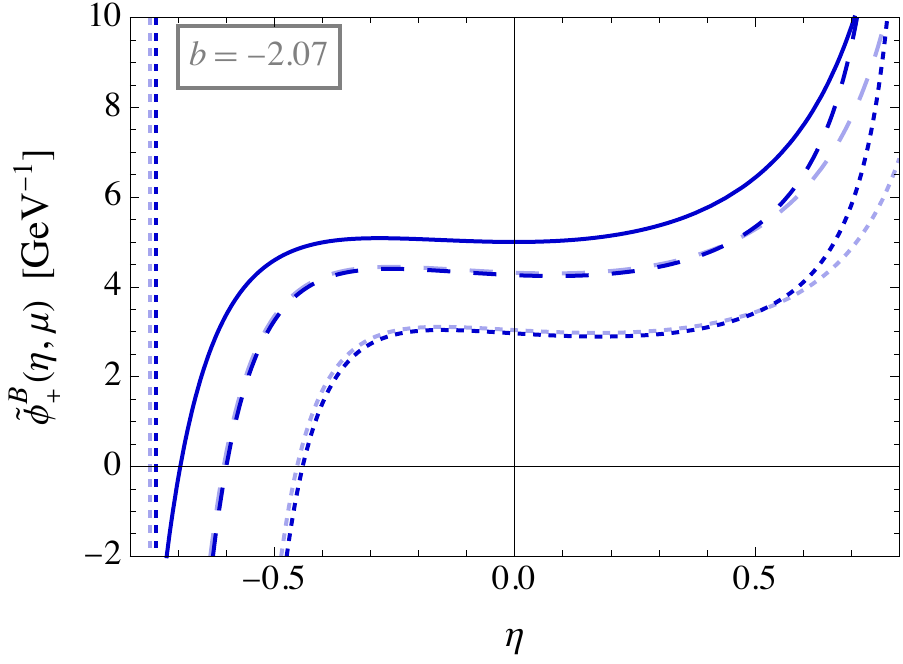}
\caption{\label{fig:modelLaplace} 
Scale evolution of the model functions for the LCDA in Laplace space with $\omega_0=482$\,MeV and $b=0$ (left), $b=-2.07$ (right). The solid lines show the results at $\mu=\mu_s=1$\,GeV, the dashed lines correspond to $\mu=1.5$\,GeV, and the dotted lines refer to the scale of the $b$-quark pole mass, $\mu=m_b=4.8$\,GeV.}
\end{figure} 

To illustrate the impact of RG evolution effects in Laplace space, we now consider the model function for the LCDA defined in \eqref{eq:modelFunction}. At the matching scale $\mu_s$, the Laplace transform of this function reads 
\begin{equation}\label{eq:NLOphitilde}
\begin{aligned}
   \tilde\phi_+^B(\eta, \mu_s)
   &= \frac{1}{\omega_0} \left( \frac{\omega_0}{\bar\omega} \right)^{-\eta} 
    \bigg[ \left(1 - \frac{b\spac(1+\eta)}{2} \right) \Gamma(1-\eta) \\
   &\hspace{2.8cm} + \frac{\alpha_s(\mu_s)}{3} \left( \cos\frac{\pi\eta}{2} \right)^{-1}
    \left( 1 - 2\ln\frac{\omega_0}{\mu_s} + \pi\tan\frac{\pi\eta}{2} \right) \bigg] \,.
\end{aligned}
\end{equation}
The singularities closest to the origin are located at $\eta=\eta_+$ and $\eta=-\eta_-$ with $\eta_\pm =1$. Using \eqref{eq:LaplaceSolu}, the RG-evolved model function at a scale $\mu>\mu_s$ can be obtained in a straightforward way. In Figure~\ref{fig:modelLaplace}, we show the scale evolution of the two model functions for the Laplace-space LCDA at NLO in RG-improved perturbation theory. The corresponding expressions for the RG functions $S_\Gamma$, $a_\Gamma$, and $a_\gamma$ are collected in Appendix~\ref{app:A}, while the expression for the integral over the function $\G$ has been given in \eqref{exponential}. For simplicity, we work with $n_f=4$ light quark flavors all the way down to the low scale $\mu=1$\,GeV, rather than matching onto a 3-flavor theory at the scale $\mu_c=m_c(m_c)\approx 1.275$\,GeV. We have checked that the effects of such a matching are numerically very small. The lines in the left (right) panel refer to the case where $b=0$ ($b=-2.07$). In each plot, the solid, dashed and dotted lines refer to $\mu=\mu_s=1$\,GeV, $\mu=1.5$\,GeV, and $\mu=m_b=4.8$\,GeV, respectively. The lines in lighter color show for comparison the results obtained at LO in RG-improved perturbation theory. At the low scale $\mu_s$ the model functions have a vanishing derivative at $\eta=0$ and turn out to be rather flat for values $|\eta|<0.3$, while they exhibit pole-type singularities at $\eta=\pm 1$. As the functions are evolved to higher scales they develop a non-zero slope at the origin, and the singularity at $\eta=-1$ is shifted toward larger values. Away from the singularities, the main effect of RG evolution is to shift the various curves downwards, corresponding to an increase in the value of $\lambda_B(\mu)$ for larger $\mu$. The impact of higher-order contributions to the evolution is most significant close to the singularities.

\subsection{\boldmath Scale evolution of $\lambda_B$ and the moments $\sigma_n^B$}
\label{sec:2.2}

The behavior of the Laplace-space LCDA $\tilde\phi_+^B(\eta,\mu)$ near the origin is governed by the moments defined in \eqref{eq:moments}. One finds
\begin{equation}
   \tilde\phi_+^B(0,\mu) = \frac{1}{\lambda_B(\mu)} \,, \qquad
   \tilde\phi_+^{B\,(n)}(0,\mu) = \frac{\sigma_n^B(\mu)}{\lambda_B(\mu)} \,,
\end{equation}
where $f^{(n)}(0,\mu)$ denotes the $n^{\rm th}$ derivative of the function $f(\eta,\mu)$ evaluated at $\eta=0$. It follows that in the vicinity of $\eta=0$, i.e.\ far away from the pole singularities, we can expand $\tilde\phi_+^B(\eta,\mu)$ in the Taylor series
\begin{equation}\label{series}
   \tilde\phi_+^B(\eta,\mu)
   \stackrel{|\eta|\ll 1}{=} \frac{1}{\lambda_B(\mu)}\,
    \bigg[ 1 + \sum_{n\ge 1}\,\frac{\eta^n}{n!}\,\sigma_n^B(\mu) \bigg] \,.
\end{equation}
If at the scale $\mu_s$ the auxiliary parameter $\bar\omega$ is chosen such that $\sigma_1^B(\mu_s)=0$, then the function $\tilde\phi_+^B(\eta,\mu_s)$ has a parabolic shape in the vicinity of the origin, with a curvature determined by $\sigma_2^B(\mu_s)$.

One can expand the evolution equation \eqref{wonderful} about $\eta=0$ to derive an infinite, coupled system of RG evolution equations for the parameter $\lambda_B$ and the logarithmic moments $\sigma_n^B$ \cite{Galda:2020epp}. The first few relations are
\begin{equation}
\begin{aligned}
   \frac{d\ln\!\lambda_B(\mu)}{d\ln\mu}
   &= \Gamma_{\mathrm{cusp}}(\alpha_s) \left[ \ln\frac{\mu}{\bar\omega} + \sigma_1^B(\mu) \right] 
    + \gamma(\alpha_s) - \G(0,\alpha_s) \,, \\
   \frac{d\sigma_1^B(\mu)}{d\ln\mu}
   &= \Gamma_{\mathrm{cusp}}(\alpha_s) \left[ \big( \sigma_1^B(\mu) \big)^2 - \sigma_2^B(\mu) \right] 
    +  \G^{(1)}(0,\alpha_s) \,, \\[1mm]
   \frac{d\sigma_2^B(\mu)}{d\ln\mu}
   &= \Gamma_{\mathrm{cusp}}(\alpha_s) \left[ \sigma_1^B(\mu)\,\sigma_2^B(\mu) - \sigma_3^B(\mu) + 4\zeta_3 \right] 
    + 2\sigma_1^B(\mu)\,\G^{(1)}(0,\alpha_s) + \G^{(2)}(0,\alpha_s) \,.
\end{aligned}
\end{equation}
Note that the RG equation for the moment $\sigma_n^B(\mu)$ involves the next higher moment $\sigma_{n+1}^B(\mu)$, and it is therefore impossible to express the solution to these equations in terms of a finite set of moments. However, given the exact solution \eqref{eq:LaplaceSolu}, it is nevertheless possible to write down the exact solution to the infinite set of couples equations in terms of the Laplace-space LCDA. We find
\begin{equation}
   \frac{1}{\lambda_B(\mu)}
   = N(\bar\omega;\mu_s,\mu)\,e^{2\gamma_E a_\Gamma}\,
    \frac{\Gamma(1+a_\Gamma)}{\Gamma(1-a_\Gamma)}\,
    \exp\!\left[\,\int\limits_{\alpha_s(\mu_s)}^{\alpha_s(\mu)}\!\frac{d\alpha}{\beta(\alpha)}\,
    \G\big(a_\Gamma(\mu_\alpha,\mu),\alpha\big) \right] \tilde\phi_+^B(a_\Gamma,\mu_s) \,,  
\end{equation}
and 
\begin{align}
   \sigma_1^B(\mu) &= \frac{\tilde\phi_+^{B\spac (1)}(a_\Gamma,\mu_s)}{\tilde\phi_+^B(a_\Gamma, \mu_s)}
    - \F(a_\Gamma) + \int\limits_{\alpha_s(\mu_s)}^{\alpha_s(\mu)}\!\frac{d\alpha}{\beta(\alpha)}\,
    \G^{(1)}\big(a_\Gamma(\mu_\alpha,\mu),\alpha\big) \,, \notag\\ 
   \sigma_2^B(\mu) 
   &= \frac{\tilde\phi_+^{B\spac (2)}(a_\Gamma,\mu_s)}{\tilde\phi_+^B(a_\Gamma, \mu_s)}
    - 2\,\frac{\tilde\phi_+^{B\spac (1)}(a_\Gamma,\mu_s)}{\tilde\phi_+^B(a_\Gamma, \mu_s)}
    \left[ \F(a_\Gamma) - \int\limits_{\alpha_s(\mu_s)}^{\alpha_s(\mu)}\!\frac{d\alpha}{\beta(\alpha)}\,
    \G^{(1)}\big(a_\Gamma(\mu_\alpha,\mu),\alpha\big) \right] \\
   &\quad + \left[ \F(a_\Gamma) - \!\int\limits_{\alpha_s(\mu_s)}^{\alpha_s(\mu)}\!\frac{d\alpha}{\beta(\alpha)}\,
    \G^{(1)}\big(a_\Gamma(\mu_\alpha,\mu),\alpha\big) \right]^2\! - \F^{(1)}(a_\Gamma) 
    + \int\limits_{\alpha_s(\mu_s)}^{\alpha_s(\mu)}\!\frac{d\alpha}{\beta(\alpha)}\,
    \G^{(2)}\big(a_\Gamma(\mu_\alpha,\mu),\alpha\big) \,, \notag
\end{align}
where $a_\Gamma\equiv a_\Gamma(\mu_s,\mu)$ for brevity. Analogous relations can be derived for the higher moments. We find it useful to define new moments by
\begin{equation}
   \lambda_n^B(\mu) = L_B^{(n)}(0,\mu) \,, \qquad \text{with} \quad
   L_B(\eta,\mu) = \ln\tilde\phi_+^B(\eta,\mu) \,, 
\end{equation}
such that close to the origin
\begin{equation}
   \phi_+^B(\eta,\mu) \stackrel{|\eta|\ll 1}{=} \frac{1}{\lambda_B(\mu)}\,
    \exp\bigg[\, \sum_{n\ge 1}\,\frac{\eta^n}{n!}\,\lambda_n^B(\mu) \bigg] \,.
\end{equation}
We then obtain the simple all-order relation (for integer $n\ge 1$)
\begin{equation}
   \lambda_n^B(\mu)
   = L_B^{(n)}(a_\Gamma,\mu_s) + \F^{(n-1)}(0) - \F^{(n-1)}(a_\Gamma) 
    + \int\limits_{\alpha_s(\mu_s)}^{\alpha_s(\mu)}\!\frac{d\alpha}{\beta(\alpha)}\,
    \G^{(n)}\big(a_\Gamma(\mu_\alpha,\mu),\alpha\big) \,,
\end{equation}
where $\F^{(n)}(0)=2\spac n!\,\zeta_{n+1}$ for even $n$, and 0 for odd $n$. In terms of the parameters $\sigma_n^B(\mu)$, we find that 
\begin{equation}
\begin{aligned}
   \sigma_1^B(\mu) &= \lambda_1^B(\mu) \,, \\
   \sigma_2^B(\mu) &= \lambda_2^B(\mu) + \left[ \lambda_1^B(\mu) \right]^2 , \\
   \sigma_3^B(\mu) 
   &= \lambda_3^B(\mu) + 3\spac\lambda_2^B(\mu)\,\lambda_1^B(\mu) + \left[ \lambda_1^B(\mu) \right]^3 , \\
   \sigma_4^B(\mu) 
   &= \lambda_4^B(\mu) + 4\spac\lambda_3^B(\mu)\,\lambda_1^B(\mu) + 3 \left[ \lambda_2^B(\mu) \right]^2
    + 6\spac\lambda_2^B(\mu) \left[ \lambda_1^B(\mu) \right]^2 
    + \left[ \lambda_1^B(\mu) \right]^4 ,
\end{aligned}
\end{equation}
and so on.

\begin{figure}[t]
\centering
\includegraphics[width=8.1cm]{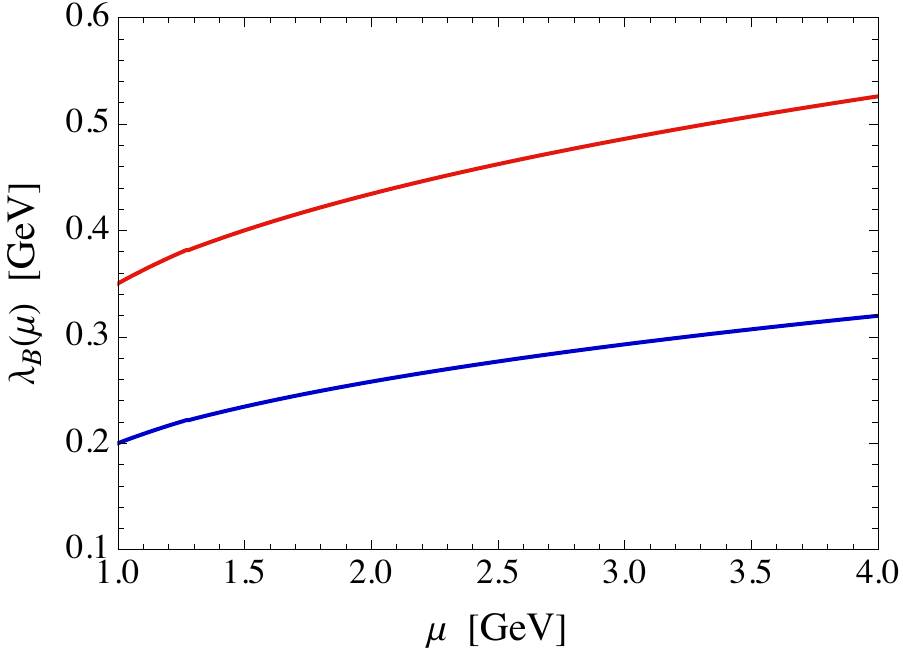}\quad
\includegraphics[width=8.1cm]{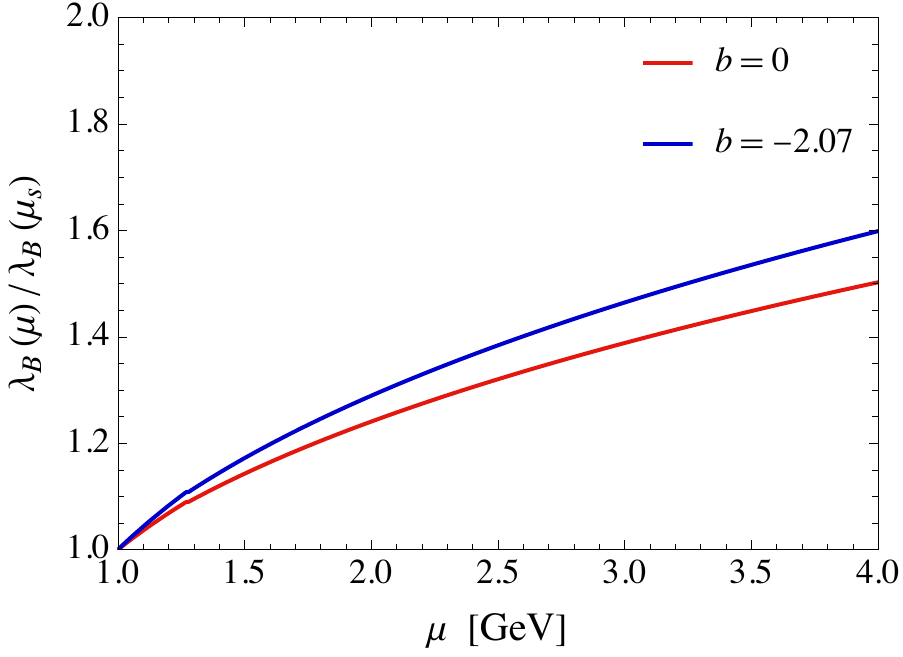}
\caption{\label{fig:lambdaB} 
Scale evolution of the hadronic parameter $\lambda_B(\mu)$ for the two model functions with $b=0$ (red curves) and $b=-2.07$ (blue curves). The small kink at $\mu=m_c(m_c)$ results from the discontinuity of $\alpha_s(\mu)$ at the threshold at which the charm quark is integrated out.}
\end{figure} 

\begin{figure}[t]
\centering
\includegraphics[width=8.4cm]{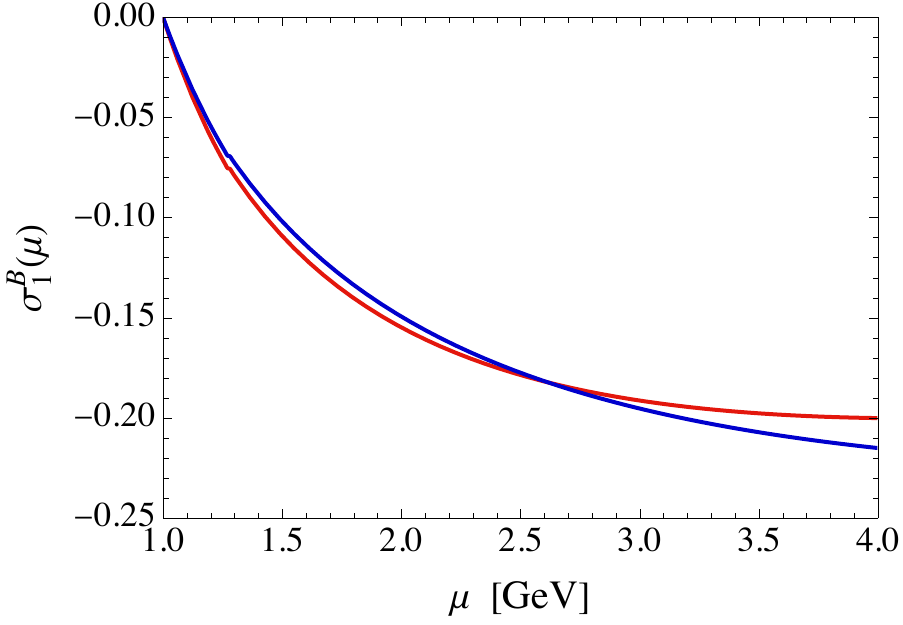}\hfill
\includegraphics[width=8.05cm]{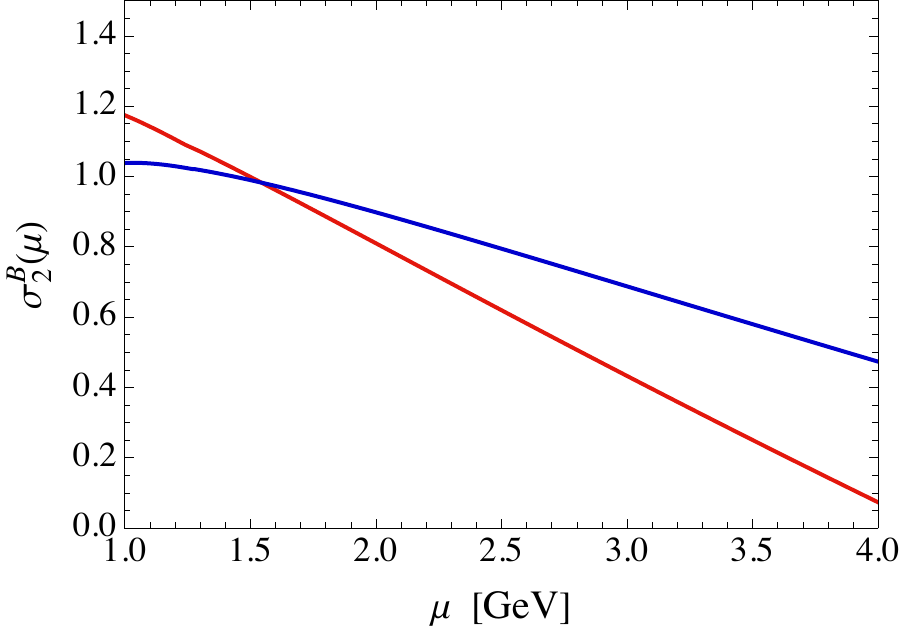} \\
\hspace{2.5mm} \includegraphics[width=8.0cm]{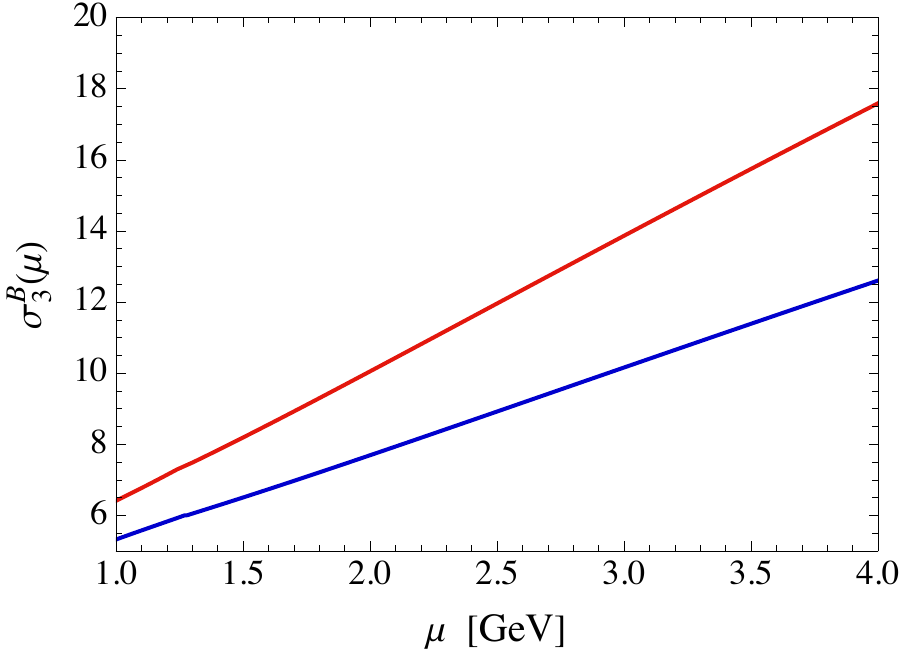}\hfill
\includegraphics[width=8.35cm]{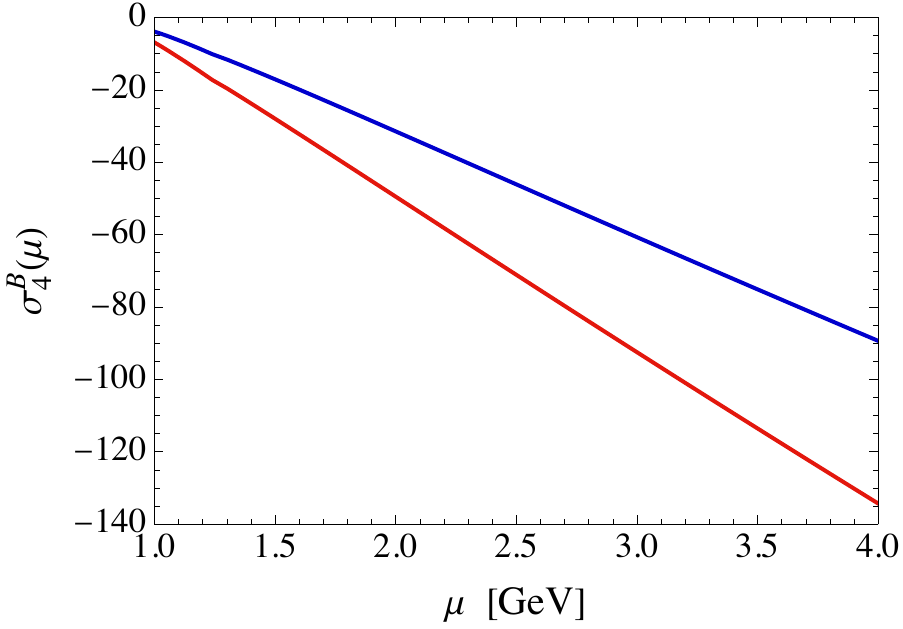}
\caption{\label{fig:moments} 
Scale evolution of the first four moments $\sigma_n^B(\mu)$ at NLO in RG-improved perturbation theory, for the two model functions with $b=0$ (red curves) and $b=-2.07$ (blue curves).}
\end{figure} 

In Figures~\ref{fig:lambdaB} and \ref{fig:moments}, we show the scale evolution of $\lambda_B$ and of the first four moments $\sigma_n^B$ for the two model functions considered in the previous section. We observe that the scale dependence of these quantities is rather significant. The values of $\lambda_B$ increase for larger $\mu$, because the LCDA broadens as the scale is increased. The relative increase is rather similar for the two model functions (right panel). The difference between the two colored curves offers a hint at the model dependence of the results. The first moment $\sigma_1^B$, which vanishes at the low scale $\mu_s=1$\,GeV by choice of $\bar\omega$, becomes negative as $\mu$ is raised to larger values. Comparing the two curves in each panel, we observe that the model dependence increases for the higher moments ($n\ge 2$). The fact that the higher moments ($n=3,4$) are larger in absolute value as the scale is increased is in line with our argument that setting $\sigma_1^B(\mu)=0$ at a given scale tends to ensure that also the higher moments take reasonably small values.

\subsection{Solution in momentum space}

Given the exact solution \eqref{eq:LaplaceSolu} of the RG equation in Laplace space, we can obtain the exact solution in momentum space by performing the inverse Laplace transformation
\begin{equation}\label{eq:inverselaplacedef}
\begin{aligned}
   \phi_+^B(\omega,\mu) 
   = \frac{1}{2\pi i} \int\limits_{c-i\infty}^{c+i\infty}\!d\eta \left( \frac{\omega}{\bar\omega} \right)^\eta
    \tilde\phi_+^B(\eta,\mu) \,.
\end{aligned}
\end{equation}
After a straightforward calculation, we obtain
\begin{equation}\label{TrafoMomentum2}
\begin{aligned}
   \phi_+^B(\omega,\mu) 
   &= N(\omega;\mu_s,\mu)\,e^{2\gamma_E\,a_\Gamma(\mu_s,\mu)} \int_0^\infty\!\frac{d\omega'}{\omega'}\,
    \phi_+^B(\omega',\mu_s)\,\frac{1}{2\pi i}\,\int\limits_{c-i\infty}^{c+i\infty}\!d\eta\,
    \Big( \frac{\omega'}{\omega} \Big)^{-\eta} \\
   &\quad \times \frac{\Gamma(1+\eta)\,\Gamma\big(1-\eta+a_\Gamma(\mu_s,\mu)\big)}%
                      {\Gamma(1-\eta)\,\Gamma\big(1+\eta-a_\Gamma(\mu_s,\mu)\big)}\,
    \exp\!\left[\, \int\limits_{\alpha_s(\mu_s)}^{\alpha_s(\mu)}\!\frac{d\alpha}{\beta(\alpha)}\,
    \G\big(\eta+a_\Gamma(\mu_\alpha,\mu_s),\alpha\big) \right] .
\end{aligned}
\end{equation}
Note that the first argument in the quantity $N$ in \eqref{eq:norm} has changed from $\bar\omega$ to $\omega$. The integration contour in the complex $\eta$-plane must be chosen to the right of the poles of $\Gamma(1+\eta)$ and to the left of the poles of $\Gamma\big(1-\eta+a_\Gamma(\mu_s,\mu)\big)$, which implies that
\begin{equation}
   -1 < c < 1 + a_\Gamma(\mu_s,\mu) \,.
\end{equation}
For $\mu>\mu_s$ we have $a_\Gamma(\mu_s,\mu)<0$, but for all realistic values of $\mu$ it is safe to assume that $a_\Gamma(\mu_s,\mu)>-1$.

While the above solution is exact, the integral over $\eta$ can in general not be evaluated in closed form. At LO in RG-improved perturbation theory, however, the function $\G$ in the exponent of the last term can be set to zero, and one obtains
\begin{equation}
   \frac{1}{2\pi i} \int\limits_{c-i\infty}^{c+i\infty}\!d\eta\,\Big( \frac{\omega'}{\omega} \Big)^{-\eta}\,
    \frac{\Gamma(1+\eta)\,\Gamma(1-\eta+a_\Gamma)}{\Gamma(1-\eta)\,\Gamma(1+\eta-a_\Gamma)} 
   = \MeijerG[\bigg]{1}{1}{2}{2}{-a_\Gamma,\,1-a_\Gamma}{1,\,0}{\frac{\omega'}{\omega}} \,,
\end{equation}
where $G$ denotes the Meijer $G$-function \cite{Gfun3}. It can be reduced to hypergeometric functions by using the theorem of residues, yielding
\begin{equation}\label{eq:MeijerGF21}
   \MeijerG[\bigg]{1}{1}{2}{2}{-a_\Gamma,\,1-a_\Gamma}{1,\,0}{z}
   = \begin{cases} 
    \frac{\Gamma(2+a_\Gamma)}{\Gamma(-a_\Gamma)}\,z\,\spac
     {}_2F_1\left(1+a_\Gamma,2+a_\Gamma;2;z\right) \,; \quad z\le 1 \,, \\[4mm]
 	\frac{\Gamma(2+a_\Gamma)}{\Gamma(-a_\Gamma)}\,z^{-1-a_\Gamma}\,\spac
	 {}_2F_1\left(1+a_\Gamma,2+a_\Gamma;2;\,\frac{1}{z}\right) \,; \quad z>1 \,.
	\end{cases}
\end{equation}
This function possesses a singularity when $z$ approaches 1. In the vicinity of the singular point one finds
\begin{equation}
   \lim_{z\to 1}\,\MeijerG[\bigg]{1}{1}{2}{2}{-a_\Gamma,\,1-a_\Gamma}{1,\,0}{z}
   = \frac{\Gamma(1+2a_\Gamma)}{\Gamma(1+a_\Gamma)\,\Gamma(-a_\Gamma)}\,
    \frac{1}{|1-z|^{1+2a_\Gamma}} + \mathcal{O}\big((1-z)^0\big) \,.
\end{equation}
Due to the fact that $a_\Gamma\equiv a_\Gamma(\mu_s,\mu)<0$ for $\mu>\mu_s$ this singularity is integrable. Implementing the asymptotic behavior shown above is particularly useful when integrating the Meijer $G$-function numerically. Combining \eqref{TrafoMomentum2} and \eqref{eq:MeijerGF21}, we recover the LO solution to the RG equation for the momentum-space LCDA found a long time ago in \cite{Lee:2005gza}, i.e.\
\begin{equation}
\begin{aligned}
   \phi_+^B(\omega,\mu) \big|_{\rm LO}
   &= \exp\big[ S_\Gamma(\mu_s,\mu) + a_\gamma(\mu_s,\mu) \big]\,
    e^{2\gamma_E\,a_\Gamma}\,\frac{\Gamma(2+a_\Gamma)}{\Gamma(-a_\Gamma)} \\
   &\quad\times \int_0^\infty\!\frac{d\omega'}{\omega'}\,\phi_+^B(\omega',\mu_s)\,
    \left( \frac{\omega_>}{\mu_s} \right)^{-a_\Gamma} \frac{\omega_<}{\omega_>}\,\,
    {}_2F_1\left(1+a_\Gamma,2+a_\Gamma;2;\frac{\omega_<}{\omega_>}\right) ,
\end{aligned}
\end{equation}
where $a_\Gamma\equiv a_\Gamma(\mu_s,\mu)$, and we have defined $\omega_<=\min(\omega,\omega')$ and $\omega_>=\max(\omega,\omega')$. In order to find the solution valid at NLO in RG-improved perturbation theory, we expand the exponential of the integral over the function $\G$ as shown in \eqref{MomentumFinal}, noting however the difference in the first argument of the $\G$ function. After a straightforward calculation, we obtain
\begin{equation}\label{MomentumFinal}
\begin{aligned}
   \phi_+^B(\omega,\mu) 
   &= N(\omega;\mu_s,\mu)\,e^{2\gamma_E\,a_\Gamma} \int_0^\infty\!\frac{d\omega'}{\omega'}\,
    \phi_+^B(\omega',\mu_s)\,\Bigg[ 
    \MeijerG[\bigg]{1}{1}{2}{2}{-a_\Gamma,\,1-a_\Gamma}{1,\,0}{\frac{\omega'}{\omega}} \\
   &\quad + \frac{C_F\spac\alpha_s(\mu_s)}{2\pi} \int_0^1\!\frac{dx}{1-x}\,\frac{h(x)}{\beta_0}\,
    \frac{1-r^{-1-\frac{2C_F}{\beta_0} \ln x}}{1+\frac{2C_F}{\beta_0} \ln x}\,
    \MeijerG[\bigg]{1}{1}{2}{2}{-a_\Gamma,\,1-a_\Gamma}{1,\,0}{\frac{x\spac\omega'}{\omega}}
    + \mathcal{O}(\alpha_s^2) \Bigg] ,
\end{aligned}
\end{equation}
where as previously $r=\alpha_s(\mu_s)/\alpha_s(\mu)$. This equation establishes the solution of the momentum-space evolution equation at NLO in RG-improved perturbation theory. It is one of the main new results of our paper.

\begin{figure}
\centering
\includegraphics[width=8.1cm]{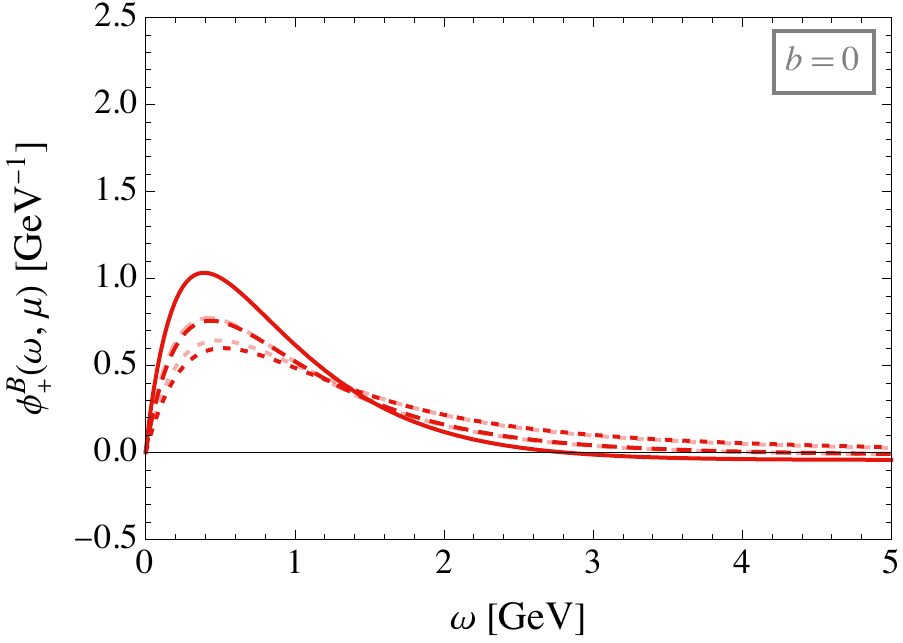}\quad
\includegraphics[width=8.1cm]{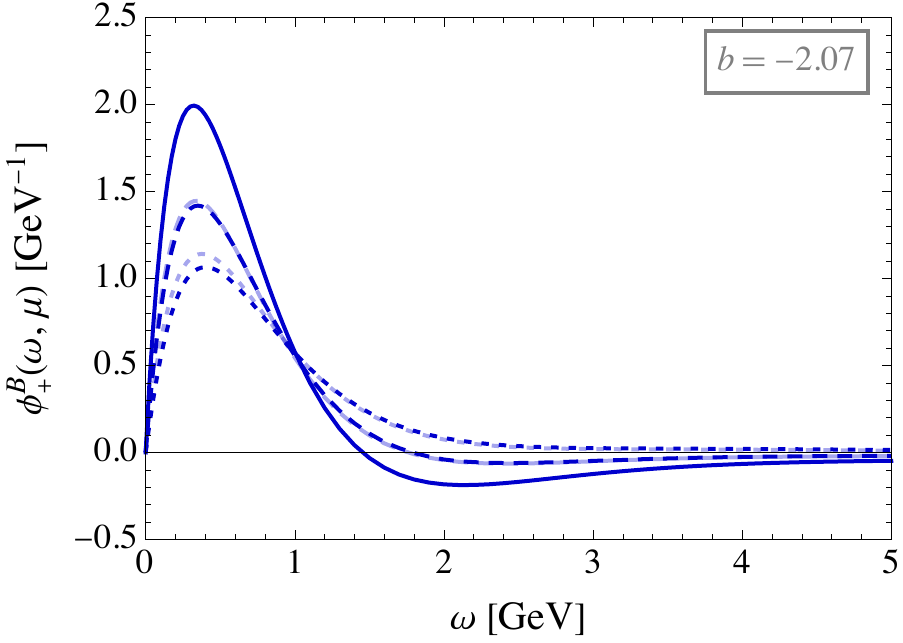}
\caption{\label{fig:momentumspace} 
Scale evolution of the model functions for the LCDA in momentum space with $\omega_0=482$\,MeV and $b=0$ (left), $b=-2.07$ (right). The solid lines show the results at $\mu=\mu_s=1$\,GeV, the dashed lines correspond to $\mu=1.5$\,GeV, and the dotted lines refer to the scale of the $b$-quark pole mass, $\mu=m_b=4.8$\,GeV.}
\end{figure}

In Figure~\ref{fig:momentumspace} we illustrate the effect of scale evolution in momentum space. The meaning of the various curves is the same as in Figure~\ref{fig:modelLaplace}. As the scale $\mu$ is increased, the LCDA is depleted in the peak region and flows toward larger $\omega$ values. The impact of higher-order contributions to the evolution is most visible in the peak region, as indicated by the lines in lighter color.

\subsection{Solution in diagonal space}
\label{sec:dualSpaceSolution}

The ``dual'' space was introduced in \cite{Bell:2013tfa,Braun:2014owa} in order to find a method that renders the one-loop RG equation for the LCDA local in the momentum variable $\omega$. The LCDA in this space, $\varphi_{\rm dual}^B(\omega,\mu)$, is related to the Laplace-space LCDA by a suitably constructed integral transformation. However, the transformation obtained in \cite{Bell:2013tfa,Braun:2014owa} no longer localizes the anomalous-dimension kernel in \eqref{anomalous} when the two-loop contribution $\hat\gamma(\omega,\omega';\alpha_s)$ is taken into account. In order to render the RG evolution equation local at two-loop order and beyond, the ``diagonal'' space was introduced in \cite{Liu:2020eqe} for the case of the soft-quark soft function, whose evolution equation shares many similarities with the RG equation of the $B$-meson LCDA. Here we apply the same method to discuss the RG evolution of the LCDA in the space where its anomalous dimension is diagonal in $\omega$ and $\omega'$.

The starting point is the observation that the Laplace-space solution \eqref{eq:LaplaceSolu} can be rearranged in the form
\begin{equation}\label{eq:splitup}
   f_{\rm diag}(\eta,\mu;\rho)
   = N(\bar\omega;\mu_s,\mu)\,e^{2\gamma_E\spac a_\Gamma(\mu_s,\mu)}\,
    f_{\rm diag}\big(\eta+a_\Gamma(\mu_s,\mu),\mu_s;\rho\big) \,,
\end{equation}
where
\begin{equation}
   f_{\rm diag}(\eta,\mu;\rho) 
   = \tilde\phi_+^B(\eta,\mu)\,\frac{\Gamma(1+\eta)}{\Gamma\big(1-\eta)}\,
    \exp\!\left[\, \int\limits_{\alpha_s(\mu)}^{\alpha_s(\rho)}\!\frac{d\alpha}{\beta(\alpha)}\,
    \G\big(\eta+a_\Gamma(\mu_\alpha,\mu),\alpha\big) \right] .
\end{equation}
Here $\rho$ is an auxiliary scale introduced to split up the integral over the function $\G$ into two integrals. The $B$-meson LCDA in the diagonal space is defined via the inverse Laplace transform of the function $f_{\rm diag}$, i.e.\
\begin{equation}\label{DiagLCDA}
   \varphi_{\rm diag}^B(\omega,\mu;\rho) 
   \equiv \frac{1}{2\pi i} \int\limits_{c-i\infty}^{c+i\infty}\!d\eta\,f_{\rm diag}(\eta,\mu;\rho)
    \left( \frac{\omega}{\bar\omega} \right)^\eta , 
\end{equation}
with a suitably chosen constant $c$. It follows from \eqref{eq:splitup} that the scale evolution of this function (for fixed choice of $\rho$) is multiplicative,
\begin{equation}\label{eq:diagsol}
   \varphi_{\rm diag}^B(\omega,\mu;\rho) 
   = N(\omega,\mu_s,\mu)\,e^{2\gamma_E\,a_\Gamma(\mu_s,\mu)}\,\varphi_{\rm diag}^B(\omega,\mu_s;\rho) \,,
\end{equation}
which is vastly simpler than the solution \eqref{TrafoMomentum2} in momentum space. In fact, relation \eqref{eq:diagsol} is the solution to the {\em local\/} (in $\omega$) evolution equation 
\begin{equation}
   \frac{d}{d\ln\mu}\,\varphi_{\rm diag}^B(\omega,\mu;\rho) 
   = - \left[ \Gamma_{\rm cusp}(\alpha_s)\,\ln\frac{\mu}{\omega\spac e^{-2\gamma_E}} + \gamma(\alpha_s) \right]
    \varphi_{\rm diag}^B(\omega,\mu;\rho) \,.
\end{equation}
In other words, the transformation \eqref{DiagLCDA} diagonalizes the non-local evolution kernel \eqref{anomalous} to all orders of perturbation theory.

The LCDAs in the diagonal space and in momentum space are related to each other via the integral transformations \cite{Liu:2020eqe}
\begin{equation}\label{integraltransforms}
\begin{aligned}
   \varphi_{\rm diag}^B(\omega,\mu;\rho) 
   &= \int_0^\infty\!\frac{dx}{\sqrt x}\,F_{\rm diag}(x,\mu;\rho)\,\phi_+^B(x\omega,\mu) \,, \\
   \phi_+^B(\omega,\mu) &= \int_0^\infty\!\frac{dx}{\sqrt x}\,F_{\rm diag}^{\rm inv}(x,\mu;\rho)\,
    \varphi_{\rm diag}^B\bigg(\frac{\omega}{x},\mu;\rho\bigg) \,,
\end{aligned}
\end{equation}
with transfer functions given by
\begin{equation}\label{FPhi}
\begin{aligned}
   \sqrt{x}\,F_{\rm diag}(x,\mu;\rho) 
   &= \frac{1}{2\pi i}\,\int\limits_{c-i\infty}^{c+i\infty}\!d\eta\,\frac{\Gamma(1+\eta)}{\Gamma(1-\eta)}\,
    x^{-\eta}\,\exp\Bigg[\,\int\limits_{\alpha_s(\mu)}^{\alpha_s(\rho)}\!\frac{d\alpha}{\beta(\alpha)}\,
    \G\big(\eta+a_\Gamma(\mu_\alpha,\mu),\alpha\big) \Bigg] \,, \\
   \sqrt{x}\,F_{\rm diag}^{\rm inv}(x,\mu;\rho) 
   &= \frac{1}{2\pi i}\,\int\limits_{c-i\infty}^{c+i\infty}\!d\eta\,\frac{\Gamma(1-\eta)}{\Gamma(1+\eta)}\,
    x^{\eta}\,\exp{\Bigg[ -\int\limits_{\alpha_s(\mu)}^{\alpha_s(\rho)}\,\frac{d\alpha}{\beta(\alpha)}\,
    \G\big(\eta+a_\Gamma(\mu_\alpha,\mu),\alpha\big) \Bigg]} \,,
\end{aligned}
\end{equation}
which obey the orthonormality condition \cite{ Liu:2020eqe}
\begin{equation}\label{Orthonomality}
   \int_0^\infty\!dx\,F_{\rm diag}(ax,\mu;\rho)\,F_{\rm diag}^{\rm inv}(bx,\mu;\rho) = \delta(a-b) \,.
\end{equation}

For practical applications of these results, it is useful to expand the transfer functions in powers of $\alpha_s(\mu)$. We obtain
\begin{equation}
   F_{\rm diag}(x,\mu;\rho) = F_{\rm diag}^{[0]}(x) + \frac{\alpha_s(\mu)}{4\pi}\,F_{\rm diag}^{[1]}(x,r_\rho) 
    + \mathcal{O}(\alpha_s^2) \,,
\end{equation}
where $r_\rho=\alpha_s(\rho)/\alpha_s(\mu)$, and similarly for the function $F_{\rm diag}^{\rm inv}$. The expansion coefficients are obtained by expanding the exponential of the integral over $\G$ in powers of $\alpha_s$. This yields
\begin{equation}
\begin{aligned}
   F_{\rm diag}^{[0]}(x) &= F_{\rm diag}^{{\rm inv}\,[0]}(x) = J_1(2\sqrt{x}) \,, \\
   F_{\rm diag}^{[1]}(x,r_\rho) &= - \frac{2C_F}{\beta_0} \int_0^1\!\frac{dy}{1-y}\,h(y)\,
    \frac{r_\rho^{1+\frac{2C_F}{\beta_0} \ln y}-1}{1+\frac{2C_F}{\beta_0} \ln y}\,\sqrt{y}\,J_1(2\sqrt{xy}) \,, \\
   F_{\rm diag}^{{\rm inv}\,[1]}(x,r_\rho) 
   &= \frac{2C_F}{\beta_0} \int_0^1\!\frac{dy}{1-y}\,h(y)\,
    \frac{r_\rho^{1+\frac{2C_F}{\beta_0} \ln y}-1}{1+\frac{2C_F}{\beta_0} \ln y}\,\sqrt{\frac{1}{y}}\,
    J_1\bigg(2\sqrt{\frac{x}{y}}\bigg) \,,
\end{aligned}
\end{equation}
where $J_1(x)$ is a Bessel function. The leading-order transfer function was first obtained in \cite{Bell:2013tfa}.

Starting at two-loop order, the construction of the diagonal space requires the introduction of the auxiliary scale $\rho$. Following \cite{Liu:2020eqe}, we find that 
\begin{equation}
   \frac{d}{d\ln\rho}\,\varphi_{\rm diag}^B(\omega,\mu;\rho) 
   = \left[ C_F \left( \frac{\alpha_s(\rho)}{2\pi} \right)^2 \int_0^1\!dx\,\frac{h(x)}{1-x}\,x^{a_\Gamma(\mu,\rho)}
    +\mathcal{O}(\alpha_s^3) \right] \varphi_{\rm diag}^B\bigg(\frac{\omega}{x},\mu;\rho\bigg) \,.
\end{equation}
The dependence on $\rho$ cancels in the product of all functions in a QCD factorization theorem. A particularly convenient choice for our purposes is to set $\rho=\mu_s$, where $\mu_s$ is the scale at which the hadronic input for the LCDA is provided. With this particular choice, one finds that
\begin{equation}\label{simplechoice}
\begin{aligned}
   \varphi_{\rm diag}^B(\omega,\mu_s;\mu_s) 
   &= \int_0^\infty\!\frac{dx}{\sqrt x}\,J_1(2\sqrt{x})\,\phi_+^B(x\omega,\mu_s) \,, \\
   \phi_+^B(\omega,\mu_s) &= \int_0^\infty\!\frac{dx}{\sqrt x}\,J_1(2\sqrt{x})\,
    \varphi_{\rm diag}^B\bigg(\frac{\omega}{x},\mu_s;\mu_s\!\bigg) \,.
\end{aligned}
\end{equation}
In practice, the diagonal space offers no advantage over the solutions in momentum space or Laplace space, because the transfer functions \eqref{FPhi} are very complicated even at LO.

\begin{figure}
\centering
\includegraphics[width=8.1cm]{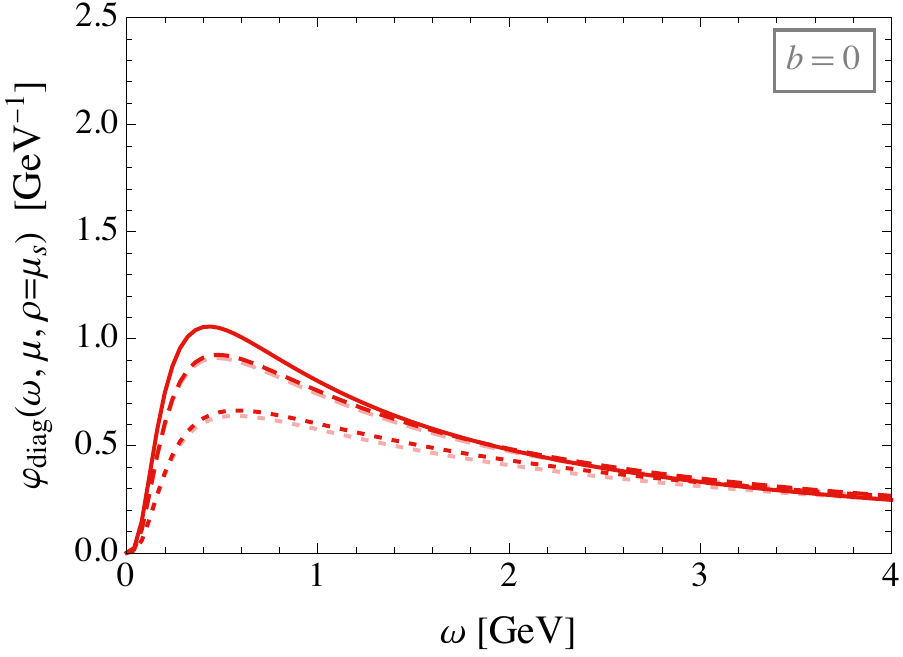}\quad
\includegraphics[width=8.1cm]{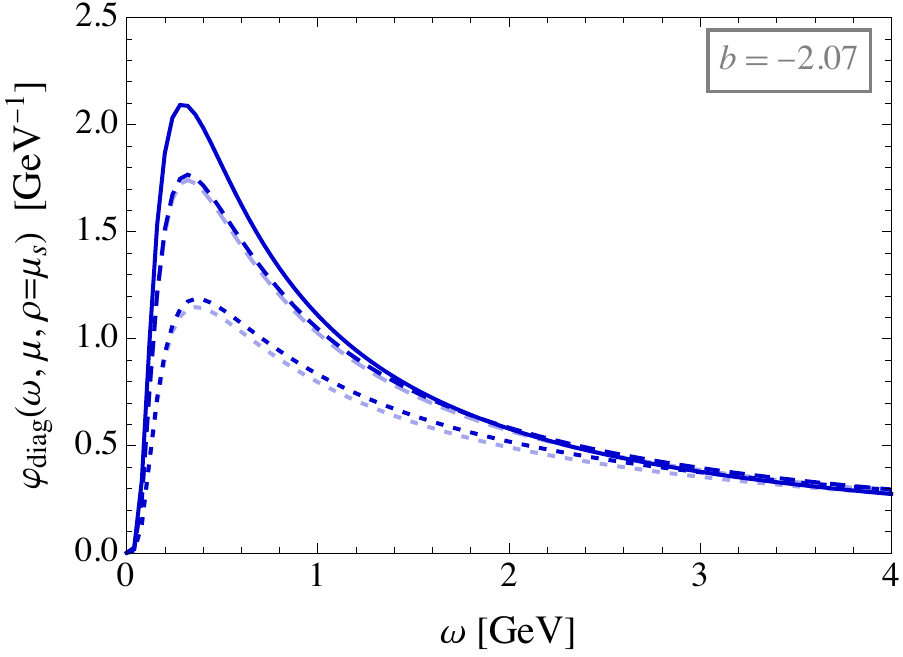}
\caption{\label{fig:Dualspace} 
Scale evolution of the model functions for the LCDA in the diagonal space with $\omega_0=482$\,MeV and $b=0$ (left), $b=-2.07$ (right). The solid lines show the results at $\mu=\mu_s=1$\,GeV, the dashed lines correspond to $\mu=1.5$\,GeV, and the dotted lines refer to the scale of the $b$-quark pole mass, $\mu=m_b=4.8$\,GeV.}
\end{figure}

We now consider the model function \eqref{eq:modelFunction} to illustrate the effects of scale evolution in diagonal space. Ignoring the radiative tail of the model functions for simplicity (we have not succeeded to obtain an analytic form of the integral transformation for this term), we obtain the simple form
\begin{equation}\label{eq:dualNLOFO}
   \varphi_{\rm diag}^B(\omega,\mu_s;\mu_s)
   = \left( \frac{1}{\omega} - \frac{b}{2}\,\frac{\omega_0}{\omega^2} \right) e^{-\omega_0/\omega}
    + {\cal O}\big(\alpha_s(\mu_s)\big) \,.
\end{equation}
Note, however, that in our numerical results the radiative tail is always included. In Figure~\ref{fig:Dualspace} we illustrate the effect of scale evolution in the diagonal space. The meaning of the various curves is the same as in Figure~\ref{fig:modelLaplace}. As the scale $\mu$ is increased, the LCDA is depleted in a more uniform way than in momentum space.

\section{Scale-invariant factorization formula}

We now return to the problem of deriving an RG-improved result for the $B^-\to\gamma\,\ell^-\spac\bar\nu$ decay amplitude in \eqref{fact}, in which all large logarithmic corrections are resummed. As shown in \cite{Galda:2020epp}, our exact solution of the evolution equation for the LCDA in Laplace space, combined with known solutions for the evolution equations of the hard and jet functions, allows one to derive a master formula in which all logarithmically enhanced terms are resummed, and which is explicitly (not only implicitly) independent of the factorization scale $\mu$. Here we provide the technical details of this calculation. The representation of the factorization formula in the diagonal space is discussed in Appendix~\ref{app:B}.

\subsection{Evolution equations}

As written in \eqref{fact}, the convolution integral 
\begin{equation}\label{eq:Idef}
   I = H(m_b,E_\gamma,\mu)\!\int_0^\infty\!\frac{d\omega}{\omega}\,
    J(-2E_\gamma\spac\omega,\mu)\,\phi_+^B(\omega,\mu) 
\end{equation}
is independent of the factorization scale $\mu$. This fact is reflected by the RG evolution equations obeyed by the hard and jet functions and by the LCDA. The evolution of the hard function is governed by the equation (here and below $\alpha_s\equiv\alpha_s(\mu)$) \cite{Bauer:2001yt,Becher:2003kh}
\begin{equation}
   \frac{d}{d\ln\mu}\,H(m_b,E_\gamma,\mu) 
   = \left[ \Gamma_{\rm cusp}(\alpha_s)\,\ln\frac{2E_\gamma}{\mu} + \gamma_H(\alpha_s) \right]  
    H(m_b,E_\gamma,\mu) \,.
\end{equation}
The cusp anomalous dimension $\Gamma_{\rm cusp}(\alpha_s)$ is known at four-loop order \cite{Henn:2019swt}, while the anomalous dimension $\gamma_H(\alpha_s)$ is known to three loops. It can be written as \cite{Becher:2009kw}
\begin{equation}
   \gamma_H(\alpha_s) = \gamma_q(\alpha_s) + \gamma_Q(\alpha_s) - \gamma_F(\alpha_s) \,,
\end{equation}
where the quantities on the right-hand side are the anomalous dimension for a light quark, a heavy quark, and the hard matching coefficient $K_F(\mu)$ in \eqref{KFdef}, which satisfies \cite{Broadhurst:1991fz}
\begin{equation}
   \frac{d}{d\ln\mu}\,K_F(m_b,\mu) = \gamma_F(\alpha_s)\,K_F(m_b,\mu) \,.
\end{equation}
The three-loop expressions for these quantities were obtained in \cite{Moch:2005id,Becher:2006mr} for $\gamma_q$, \cite{Bruser:2019yjk} for $\gamma_Q$, and \cite{Chetyrkin:2003vi} for $\gamma_F$. The general solution to the RG evolution equation for the hard function can be written in the form
\begin{equation}\label{eq:Hevol}
   H(m_b,E_\gamma,\mu)
   = H(m_b,E_\gamma,\mu_h) \left(\frac{2E_\gamma}{\mu_h} \right)^{-a_\Gamma(\mu_h,\mu)}
    \exp\Big[S_\Gamma(\mu_h,\mu)-a_{\gamma_H}(\mu_h,\mu)\Big] \,,
\end{equation}
where $a_{\gamma_H}$ is defined in analogy with $a_\Gamma$ in \eqref{eq:aGamma}. Here $\mu_h\sim m_b$ denotes a hard matching scale, at which the initial condition for the hard function is free of large logarithms and hence can be calculated in fixed-order perturbation theory. 

The RG equation for the jet function reads \cite{Liu:2020ydl}
\begin{equation}\label{RGE}
   \frac{d}{d\ln\mu}\,J(p^2,\mu) 
   = - \int_0^\infty\!dx\,\gamma_J(p^2,x p^2;\mu)\,J(x p^2,\mu) \,.
\end{equation}
The anomalous-dimension kernel is given by 
\begin{equation}\label{gammaJ}
   \gamma_J(p^2,x p^2;\mu) 
   = \left[ \Gamma_{\rm cusp}(\alpha_s)\,\ln\frac{-p^2}{\mu^2}
    - \gamma'(\alpha_s) \right] \delta(1-x) + \Gamma_{\rm cusp}(\alpha_s)\,\Gamma(1,x) 
    + \frac{\hat\gamma(x,1;\alpha_s)}{x} \,,
\end{equation}
with the same function $\hat\gamma$ as in \eqref{anomalous}. Using these equations along with the RG equation of the LCDA shown in \eqref{RGEphiB}, it is straightforward to show that the convolution integral $I$ is scale independent if, to all orders of perturbation theory, 
\begin{equation}\label{gammarelation}
   \gamma'(\alpha_s) = \gamma(\alpha_s) - \gamma_H(\alpha_s) \,.
\end{equation}
The anomalous dimensions $\gamma(\alpha_s)$ and the kernel $\hat\gamma(\omega,\omega';\alpha_s)$ were obtained at two-loop order in \cite{Braun:2019wyx}. The general solution of the evolution equation \eqref{RGE} is \cite{Liu:2020ydl}
\begin{equation}\label{eq:Jevol}
\begin{aligned}
   J(p^2,\mu)
   &= \exp\bigg[ - 2 S_\Gamma(\mu_j,\mu) - a_{\gamma'}(\mu_j,\mu) 
    - 2\gamma_E\,a_\Gamma(\mu_j,\mu) \bigg]\,\J(\partial_\eta,\mu_j) 
    \left( \frac{-p^2-i0}{\mu_j^2} \right)^{\eta+a_\Gamma(\mu_j,\mu)} \\
   &\quad\times  
    \frac{\Gamma\big(1-\eta-a_\Gamma(\mu_j,\mu)\big)\,\Gamma(1+\eta)}%
         {\Gamma\big(1+\eta+a_\Gamma(\mu_j,\mu)\big)\,\Gamma(1-\eta)}\,
    \exp\Bigg[ - \int\limits_{\alpha_s(\mu_j)}^{\alpha_s(\mu)}\!\frac{d\alpha}{\beta(\alpha)}\, 
    \G\big(\eta+a_\Gamma(\mu_j,\mu_\alpha),\alpha\big) \Bigg] \Bigg|_{\eta=0} \,.
\end{aligned}
\end{equation}
Here $\mu_j$ is a matching scale set by the typical value of $|p^2|^{1/2}$ in a given process. At this scale, one defines $J(p^2,\mu_j)\equiv\J(L_p,\mu_j)$, where $L_p=\ln\big[(-p^2-i0)/\mu_j^2\big]$. In the solution above, the first argument in the function $\J$ is replaced by a derivative with respect to the auxiliary parameter $\eta$. The function $\G$ has been defined in \eqref{eq:eigenfunctions}, and moreover relation \eqref{gammarelation} implies that $a_{\gamma'}(\mu_j,\mu)=a_\gamma(\mu_j,\mu)-a_{\gamma_H}(\mu_j,\mu)$.

For the convenience of the reader, we list the perturbative expansion coefficients of all relevant anomalous-dimension coefficients in Appendix~\ref{app:A}. 

\subsection{Matching conditions}
\label{sec:3.2}

To complete the solutions, one needs to specify the hard and jet functions at their respective matching scales $\mu_h$ and $\mu_j$, where they can be calculated in fixed-order perturbation theory. The hard function $H=C_1/K_F$ is given by the ratio of two quantities. The first one is the short-distance Wilson coefficient $C_1(m_b,2E_\gamma,\mu)$ in the matching relation for the heavy-light current operator, 
\begin{equation}
   \bar u\spac\gamma^\mu(1-\gamma_5)\spac b
   = \sum_{i=1}^3\,C_i(m_b,2E,\mu)\,\bar\X_n\spac\Gamma_i\spac h_v 
    + {\cal O}\bigg(\frac{\Lambda_{\rm QCD}}{m_b},\frac{\Lambda_{\rm QCD}}{2E}\bigg) \,,
\end{equation}
where $h_v$ denotes the effective $b$-quark field in HQET, $\X_n=W_n^\dagger\,\xi_n$ is the effective collinear up-quark field in soft-collinear effective theory \cite{Bauer:2001yt,Bauer:2000yr}, $E\approx m_b/2$ denotes the large energy carried by the collinear quark in the rest frame of the $B$ meson, and the relevant Dirac structures can be chosen as $\Gamma_1=\gamma^\mu(1-\gamma_5)$, $\Gamma_2=v^\mu(1+\gamma_5)$ and $\Gamma_3=n^\mu(1+\gamma_5)$. The two-loop result for the coefficients $C_i$ have been obtained independently by four groups \cite{Bonciani:2008wf,Asatrian:2008uk,Beneke:2008ei,Bell:2008ws}. The second quantity entering the expression for the hard function is the matching coefficient $K_F(m_b,\mu)$ defined in relation \eqref{KFdef}, which has been calculated at two-loop order in \cite{Broadhurst:1994se,Grozin:1998kf}. Combining these result, we obtain 
\begin{equation}
\begin{aligned}
   H(m_b,E_\gamma,\mu_h)
   &= 1 + \frac{C_F\spac\alpha_s(\mu_h)}{4\pi}\,\bigg[
    - 2 L_m^2 - (2-4\ln x)\,L_m - 2\ln^2x + \frac{2-3x}{1-x} \ln x \\
   &\hspace{3.4cm} - 2\text{Li}_2(1-x) - 4 - \frac{\pi^2}{12} \bigg] \\
   &\quad + \left[ \frac{\alpha_s(\mu_h)}{4\pi} \right]^2\!\Big[
    h_0(L_m) + (1-x)\,h_1(L_m) + (1-x)^2\,h_2(L_m,x) + \dots \Big]
   + {\cal O}(\alpha_s^3) \,,
\end{aligned}
\end{equation}
where $x=2E_\gamma/m_b$ and $L_m=\ln(\mu_h/m_b)$, with $m_b=4.8$\,GeV the $b$-quark pole mass. The analytical expression for the two-loop correction in terms of harmonic polylogarithms is very lengthy. For convenience, we present here the first few terms in a Taylor expansion around $x=1$, where the numerical coefficients are obtained for $N_c=3$ colors and $n_f=4$ light (massless) quark flavors. We find
\begin{align}
   h_0(L_m)
   &= \frac{32}{9}\,L_m^4 - \frac{208}{27}\,L_m^3 + \left( - \frac{1504}{27} + \frac{80\pi^2}{27} \right) L_m^2 
    + \left( - \frac{10480}{81} + \frac{32\pi^2}{27} + \frac{136\zeta_3}{3} \right) L_m \notag\\
   &\quad - \frac{29914}{243} - \frac{1055\pi^2}{54} + \frac{940\zeta_3}{27} + \frac{856\pi^4}{405} 
    - \frac{4\pi^2}{3}\,\ln 2 \,, \notag\\
   h_1(L_m)
   &= \frac{128}{9}\,L_m^3 - \frac{128}{9}\,L_m^2 
    + \left( - \frac{4268}{27} + \frac{160\pi^2}{27} \right) L_m \notag\\
   &\quad - \frac{2422}{27} - \frac{3572\pi^2}{27} + \frac{364\zeta_3}{9} + \frac{5177\pi^4}{405} 
   - \frac{16\pi^2}{9}\,\ln 2 \,, \notag\\
   h_2(L_m,x)
   &= \frac{64}{9}\,L_m^3 + \frac{328}{27}\,L_m^2 
    + \left( - \frac{1984}{27} + \frac{80\pi^2}{27} \right) L_m 
    - \frac{193}{9}\,\ln(1-x) + \frac{193}{27}\,\ln^2(1-x) \notag\\
   &\quad + \frac{57071}{243} - \frac{61237\pi^2}{162} + \frac{11\zeta_3}{9} + \frac{4754\pi^4}{135} 
   + \frac{2\pi^2}{9}\,\ln 2 \,.
\end{align}

The jet function $J(p^2,\mu)$ has been calculated at two-loop order in \cite{Liu:2020ydl}. In our case the characteristic value of the momentum transfer are such that $p^2=-2E_\gamma\spac\omega={\cal O}(\Lambda_{\rm QCD}\spac m_b)$, since $2E_\gamma\approx m_b$ and the characteristic values of $\omega$ are governed by nonperturbative QCD dynamics. At a matching scale $\mu_j^2={\cal O}(\Lambda_{\rm QCD}\spac m_b)$, the function $\J(L_p,\mu_j)$ reads 
\begin{equation}
   \J(L_p,\mu_j) = 1 + \frac{C_F\spac\alpha_s(\mu_j)}{4\pi} \left[ \bigg( L_p^2 - 1 - \frac{\pi^2}{6} \bigg) 
   + \frac{\alpha_s(\mu_j)}{4\pi}\,\big( C_F\spac k_F + C_A\spac k_A + T_F\,n_f\spac k_{n_f} \big) \right] 
   + {\cal O}(\alpha_s^3) \,,
\end{equation}
with
\begin{equation}
\begin{aligned}
   k_F &= \frac{L_p^4}{2} - \left( 1 + \frac{\pi^2}{6} \right) L_p^2 
    + \left( \pi^2 + 4\zeta_3 \right) L_p + \frac32 - \frac{\pi^2}{3} - 39\zeta_3
    + \frac{119\pi^4}{360} \,, \\
   k_A &= - \frac{11}{9}\,L_p^3 + \left( \frac{67}{9} - \frac{\pi^2}{3} \right) L_p^2 
    - \left( \frac{305}{27} - 14\zeta_3 \right) L_p + \frac{493}{162} 
    - \frac{103\pi^2}{108} + \frac{140\zeta_3}{9} - \frac{43\pi^4}{180} \,, \\
   k_{n_f} &= \frac{4}{9}\,L_p^3 - \frac{20}{9}\,L_p^2 + \frac{76}{27}\,L_p 
    + \frac{14}{81} + \frac{5\pi^2}{27} + \frac{8\zeta_3}{9} \,.
\end{aligned}
\end{equation}

\subsection{Master formula for the convolution integral}

When the RG-improved expression for the jet function is inserted in the convolution integral in \eqref{fact}, one encounters the integral
\begin{equation}
   \int_0^\infty\!\frac{d\omega}{\omega} 
    \left( \frac{2E_\gamma\spac\omega}{\mu_j^2} \right)^{\eta+a_\Gamma(\mu_j,\mu)} \phi_+^B(\omega,\mu) 
   = \left( \frac{2E_\gamma\spac\bar\omega}{\mu_j^2} \right)^{\eta+a_\Gamma(\mu_j,\mu)} 
    \tilde\phi_+^B\big(-\eta-a_\Gamma(\mu_j,\mu),\mu\big) \,,
\end{equation}
which is given in terms of the Laplace-space LCDA defined in \eqref{eq:Laplacedef}. We now combine the relations \eqref{eq:Hevol}, \eqref{eq:Jevol} and \eqref{eq:LaplaceSolu} to derive a RG-improved expression for the decay amplitude. Using then the identities 
\begin{equation}
\begin{aligned}
   a_\Gamma(\nu_1,\mu) - a_\Gamma(\nu_2,\mu) 
   &= a_\Gamma(\nu_1,\nu_2) \,, \\
   S_\Gamma(\nu_1,\mu) - S_\Gamma(\nu_2,\mu) 
   &= S_\Gamma(\nu_1,\nu_2) - a_\Gamma(\nu_2,\mu)\,\ln\frac{\nu_1}{\nu_2} \,,
\end{aligned}
\end{equation}
which follow from the definitions \eqref{eq:aGamma} and \eqref{eq:RGFunctions}, one finds that all reference to the factorization scale $\mu$ cancels in the final expression. This leads to the master formula \cite{Galda:2020epp}
\begin{equation}\label{master}
\begin{aligned}
   I &= H(m_b,E_\gamma,\mu)\!\int_0^\infty\!\frac{d\omega}{\omega}\,
    J(-2E_\gamma\spac\omega,\mu)\,\phi_+^B(\omega,\mu) \\
   &= \exp\Big[ S_\Gamma(\mu_h,\mu_j) + S_\Gamma(\mu_s,\mu_j) - a_{\gamma_H}(\mu_h,\mu_j) 
    + a_\gamma(\mu_s,\mu_j) + 2\gamma_E\spac a_\Gamma(\mu_s,\mu_j) \Big] \\
   &\quad\times H(m_b,E_\gamma,\mu_h) \left( \frac{2E_\gamma}{\mu_h} \right)^{-a_\Gamma(\mu_h,\mu_j)}
    \J(\partial_\eta,\mu_j) \left( \frac{2E_\gamma\spac\bar\omega}{\mu_j^2} \right)^\eta
    \frac{\Gamma\big(1-\eta+a_\Gamma(\mu_s,\mu_j)\big)\,\Gamma(1+\eta)}%
         {\Gamma\big(1+\eta-a_\Gamma(\mu_s,\mu_j)\big)\,\Gamma(1-\eta)} \\[-1mm]
   &\quad\times \exp\Bigg[\,\int\limits_{\alpha_s(\mu_s)}^{\alpha_s(\mu_j)}\!\frac{d\alpha}{\beta(\alpha)}\,
    \G\big(-\eta+a_\Gamma(\mu_\alpha,\mu_j),\alpha\big) \Bigg]
    \left( \frac{\bar\omega}{\mu_s} \right)^{-a_\Gamma(\mu_s,\mu_j)} 
    \tilde\phi_+\big(\!-\!\eta+a_\Gamma(\mu_s,\mu_j),\mu_s\big) \bigg|_{\eta=0} \spac .
\end{aligned}
\end{equation}
In this expression, all large logarithmic corrections are resummed in the RG coefficients $S_\Gamma$ and $a_\Gamma$, $a_{\gamma_H}$, $a_\gamma$. The result depends on the three matching scales $\mu_h\sim m_b$, $\mu_j\sim\sqrt{\Lambda_{\rm QCD}\spac m_b}$, and the low scale $\mu_s$, at which the model function for the LCDA is assumed. However, at any given order in the RG-improved perturbation theory this dependence cancels out up to higher-order corrections.

In our numerical analysis below, we will find that the RG evolution effects from the low scale $\mu_s$ to the intermediate scale $\mu_j$ are necessarily very small. The reason is that $\mu_s$ cannot be chosen smaller than about 1\,GeV, since it needs to be in the perturbative regime. On the other hand, the master formula indicates that $\mu_j$ should be chosen such that $\mu_j^2\approx 2E_\gamma\spac\bar\omega\approx m_b\spac\bar\omega$, and with $\bar\omega$ values such as those shown in the table following equation \eqref{eq:modelFunction} one finds $\mu_j\approx 1$\,GeV. Even for a larger value such as $\mu_j=1.5$\,GeV, one obtains $a_\Gamma(\mu_s,\mu_j)\approx-0.098$, which is a small effect. This implies that the first argument of the Laplace-space LCDA is close to the origin, and hence the Taylor series in \eqref{series} can be used to obtain a model-independent parameterization of the LCDA in terms of the parameters $\lambda_B$, $\bar\omega$ and $\sigma_n^B$ with $n\ge 2$, all defined at the low scale $\mu_s$.

In light of the above remarks, one may even consider setting the two scales $\mu_j$ and $\mu_s$ equal to each other. This leads to the much simpler formula
\begin{equation}
\begin{aligned}
   I &= \exp\Big[ S_\Gamma(\mu_h,\mu_s) - a_{\gamma_H}(\mu_h,\mu_s) \Big]\,
    H(m_b,E_\gamma,\mu_h) \left( \frac{2E_\gamma}{\mu_h} \right)^{-a_\Gamma(\mu_h,\mu_s)} \\
   &\quad\times \J(\partial_\eta,\mu_s) \left( \frac{2E_\gamma\spac\bar\omega}{\mu_s^2} \right)^\eta
    \tilde\phi_+(-\eta,\mu_s) \bigg|_{\eta=0} \,.
\end{aligned}
\end{equation}

\subsection{Numerical results}

We are now ready to present our numerical results for the convolution integral in the master formula \eqref{master}, which governs the $B^-\to\gamma\,\ell^-\spac\bar\nu$ decay amplitude at leading order in the expansion in powers of $\Lambda_{\rm QCD}/m_b$, following the discussion presented in \cite{Galda:2020epp}. Based on known results for the relevant anomalous dimensions and matching conditions, we can evaluate the convolution integral at (approximate) NNLO in RG-improved perturbation theory. This requires the two-loop expressions for the hard function $H(m_b,E_\gamma,\mu_h)$ and the jet function $\J(L_p,\mu_j)$ given in Section~\ref{sec:3.2}, the four-loop expression for the cusp anomalous dimension (needed for the calculation of the Sudakov exponent $S_\Gamma$), and three-loop expressions for the remaining anomalous dimensions (needed for the calculation of the exponents $a_\Gamma$, $a_{\gamma_H}$, $a_\gamma$ and the function $\G$). At present one can only achieve approximate NNLO accuracy, since the anomalous dimension $\gamma$ in \eqref{anomalous} and the function $\G$ in \eqref{eq:eigenfunctions} are only known at two-loop order. However, in practice this is not a limitation, because the scales $\mu_s$ and $\mu_j$ are rather close to each other, and evolution effects between these scales have only a minor impact. In fact, Figure~\ref{fig:modelLaplace} shows that in the vicinity of the origin even the effects of NLO scale evolution are hardly visible. Note that for the special scale choice $\mu_j=\mu_s$ our predictions have strict NNLO accuracy.

When we derive the perturbative expansions in RG-improved perturbation theory, we consistently expand out higher-order terms in $\alpha_s$ in a perturbative series. We will denote the result as $I_{\rm aNNLO}$. Alternatively, one could perform the expansion of the RG functions $S_\Gamma$ and $a_\Gamma$, $a_{\gamma_H}$, $a_\gamma$ in the exponent (denoted by $I_{\rm aNNLO}'$). Both approximations have the same parametric accuracy, but the differences between the results obtained in these two ways can serve as an estimator of unknown higher-order corrections (see below). For comparison, we will also show results obtained at NLO in RG-improved perturbation theory.

In order to present our results we fix the photon energy (defined in the rest frame of the $B$ meson) to 2.2\,GeV and vary $\mu_h$ by a factor of two around its default value of $\mu_h=m_b=4.8$\,GeV. In addition, we fix $\mu_s=1.0$\,GeV and vary $\mu_j$ by a factor of $\sqrt{2}$ around its default value of $\sqrt{2}$\,GeV, as proposed in \cite{Galda:2020epp}. Finally, we set the parameter $\bar\omega$ equal to the reference value 300\,GeV. The dependence on this choice will be investigated later. To good approximation, one finds that changing the value of $\bar\omega$ has the effect of changing the convolution integral by a factor $(\bar\omega/300\,\text{MeV})^{-a_\Gamma(\mu_s,\mu_j)}$, where the exponent is approximately 0.1 for our default scale choices. 

With these parameter and scale choices, we obtain
\begin{equation}\label{numbers}
\begin{aligned}
   I_{\rm aNNLO} 
   &= \frac{1}{\lambda_B}\,\Big[ \big( 0.664\,_{-0.013}^{+0.011}\,_{-0.038}^{+0.024} \big) 
    + \big( 4.36\,_{-0.08}^{+0.15}\,_{-0.45}^{+0.07} \big)\cdot 10^{-2}\,\sigma_2^B \\
   &\hspace{1.3cm} + \big( 0.35\,_{-0.02}^{+0.12}\,_{-1.99}^{+2.97} \big)\cdot 10^{-3}\,\sigma_3^B 
    + \big( 5.02\,_{-0.08}^{+0.28}\,_{-1.91}^{+5.84} \big)\cdot 10^{-4}\,\sigma_4^B + \dots \Big] \,,
\end{aligned}
\end{equation}
where for each value the quoted errors arise from the variations of $\mu_h$ and $\mu_j$. In this expression, the hadronic parameters $\lambda_B$, $\sigma_n^B$ and $\bar\omega$ are defined at the reference scale $\mu_s=1$\,GeV. We observe that the uncertainties from scale variation are rather small for $\mu_h$, but significantly larger for $\mu_j$, especially as far as the coefficients of the higher moments are concerned. The moment expansion itself appears to be well behaved. If instead the perturbative expansion of the RG coefficients is performed in the exponent, one finds
\begin{equation}
\begin{aligned}
   I_{\rm aNNLO}' 
   &= \frac{1}{\lambda_B}\,\Big[ \big( 0.675\,_{-0.013}^{+0.014}\,_{-0.040}^{+0.022} \big) 
    + \big( 4.14\,_{-0.08}^{+0.08}\,_{-0.48}^{+0.12} \big)\cdot 10^{-2}\,\sigma_2^B \\
   &\hspace{1.3cm} + \big( -0.14\,_{-0.003}^{+0.003}\,_{-1.85}^{+3.00} \big)\cdot 10^{-3}\,\sigma_3^B 
    + \big( 4.33\,_{-0.08}^{+0.09}\,_{-1.42}^{+5.02} \big) \cdot 10^{-4}\,\sigma_4^B + \dots \Big] \,.
\end{aligned}
\end{equation}
Comparison with \eqref{numbers} shows that the two results are consistent with each other within the quoted errors. In order to study the impact of the NNLO corrections, it is instructive to compare our result with the one obtained at NLO in RG-improved perturbation theory. It reads 
\begin{equation}
\begin{aligned}
   I_{\rm NLO} 
   &= \frac{1}{\lambda_B}\,\Big[ \big( 0.731\,_{-0.014}^{+0.015}\,_{-0.047}^{+0.019} \big) 
    + \big( 3.53\,_{-0.06}^{+0.21}\,_{-0.76}^{+0.93} \big)\cdot 10^{-2}\,\sigma_2^B \\
   &\hspace{1.3cm} + \big( -2.75\,_{-0.17}^{+0.05}\,_{-0.91}^{+2.75} \big)\cdot 10^{-3}\,\sigma_3^B 
    + \big( 1.09\,_{-0.02}^{+0.07}\,_{-1.09}^{+1.43} \big) \cdot 10^{-4}\,\sigma_4^B + \dots \Big] \,.
\end{aligned}
\end{equation}
The central value of the leading term is about 10\% larger than at aNNLO, indicating that the higher-order effects are indeed significant and should be included in phenomenological analyses of the $B^-\to\gamma\,\ell^-\bar\nu$ photon spectrum. Within the quoted errors, the two values are nevertheless consistent with each other.

\begin{figure}
\begin{center}
\vspace{1.3mm}
\includegraphics[width=0.5\textwidth]{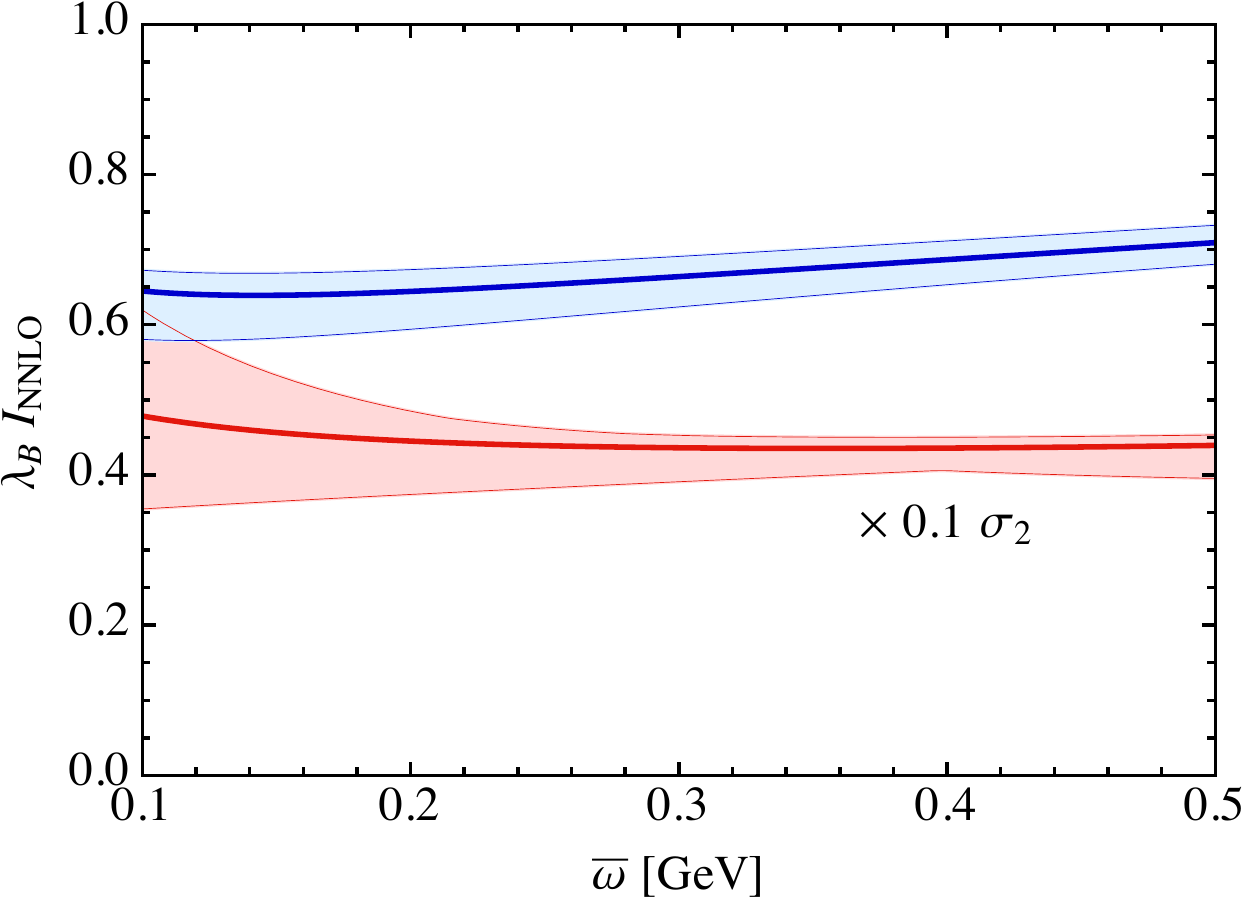}
\end{center}
\vspace{-5mm}
\caption{\label{fig:wbar} 
Coefficients of the leading term (blue) and of $0.1\sigma_2$ (red) in the result (\ref{numbers}) including scale variations added in quadrature, for different values of $\bar\omega$.}
\end{figure}

The result (\ref{numbers}) refers to $\bar\omega=300$\,MeV. Figure~\ref{fig:wbar} shows how the coefficients of the leading term and of $\sigma_2$ vary with $\bar\omega$. The range shown is motivated by the fact that parametrically $\bar\omega\sim\Lambda_{\rm QCD}$. The leading coefficient increases slightly with $\bar\omega$, whereas the coefficient of $\sigma_2$ is almost independent of it. Note that the scale variations increase for smaller values of $\bar\omega$. As can be seen from (\ref{master}), the quantity $2E_\gamma\spac\bar\omega$ sets the ``natural'' scale for $\mu_j^2$, and for $\bar\omega<0.23$\,GeV this scale drops below 1\,GeV, outside the range of variation of $\mu_j$. This suggests that the perturbative corrections to the jet function get larger the smaller $\bar\omega$ is. 

All results shown above refer to the reference choice $\mu_s=1$\,GeV. One is, of course, free to make a different choice $\mu_s'$. It is important to realize that in this case the parameters $\bar\omega$, $\lambda_B$ and $\sigma_n^B$ have a different meaning, because they refer to a different LCDA, obtained form the previous one by RG evolution. For illustration, we present our result for the case of a higher reference scale $\mu_s'=1.5$\,GeV, in this case varying the intermediate scale $\mu_j$ by a factor of $\sqrt{2}$ about the default value $\mu_j=\mu_s=1.5$\,GeV. As previously we take $\bar\omega=300$\,MeV as our reference value. In this way, we obtain 
\begin{equation}\label{resunew}
\begin{aligned}
   I_{\rm aNNLO} 
   &= \frac{1}{\lambda_B}\,\Big[ \big( 0.712_{-0.014}^{+0.011}\,_{-0.041}^{+0.029} \big)
    + \big( 4.24_{-0.08}^{+0.14}\,_{-0.40}^{+0.02} \big)\times 10^{-2}\,\sigma_2^B \\
   &\hspace{1.3cm} + \big( 4.22_{-0.07}^{+0.26}\,_{-2.43}^{+2.68} \big)\times 10^{-3}\,\sigma_3^B
    + \big( 6.69_{-0.12}^{+0.42}\,_{-3.81}^{+7.91} \big)\times 10^{-4}\,\sigma_4^B + \dots \Big] \,.
\end{aligned}
\end{equation}
Let us work out how the parameters $\lambda_B$ and $\sigma_n^B$ in this result are related to the parameters in \eqref{numbers}. When the LCDA is evolved from the scale $\mu_s$ to a different scale $\mu_s'$ (at fixed $\bar\omega$), the values of $\lambda_B(\mu_s)$ and $\sigma_n^B(\mu_s)$ evolve to new values $\lambda_B(\mu_s')$ and $\sigma_n^B(\mu_s')$, as discussed in detail in Section~\ref{sec:2.2}. In general, this leads to a non-zero first moment $\sigma_1^B(\mu_s')\ne 0$. We must now readjust the parameter $\bar\omega$ such that $[\sigma_1^B(\mu_s')]_{\rm new}=0$. According to \eqref{eq:moments}, this leads to $\bar\omega_{\rm new}=\bar\omega\,e^{-\sigma_1^B(\mu_s')}$. With this new reference scale, one finds the new moments 
\begin{equation}
\begin{aligned}
   \left[ \lambda_B(\mu_s') \right]_{\rm new}
   &= \lambda_B(\mu_s') \,, \\
   \left[ \sigma_2^B(\mu_s') \right]_{\rm new}
   &= \sigma_2^B(\mu_s') - \left[ \sigma_1^B(\mu_s') \right]^2 , \\
   \left[ \sigma_3^B(\mu_s') \right]_{\rm new}
   &= \sigma_3^B(\mu_s') - 3\spac\sigma_2^B(\mu_s')\,\sigma_1^B(\mu_s')
   + 2 \left[ \sigma_1^B(\mu_s') \right]^3 , \\
   \left[ \sigma_4^B(\mu_s') \right]_{\rm new}
   &= \sigma_4^B(\mu_s') - 4\spac\sigma_3^B(\mu_s')\,\sigma_1^B(\mu_s')
   + 6\spac\sigma_2^B(\mu_s') \left[ \sigma_1^B(\mu_s') \right]^2 - 3 \left[ \sigma_1^B(\mu_s') \right]^4 ,
\end{aligned}
\end{equation}
etc., and these are the parameters entering the result \eqref{resunew}. Note that, if $\mu_s'>\mu_s$, the first moment $\sigma_1^B(\mu_s')$ is negative, and hence $\bar\omega_{\rm new}>\bar\omega$ in this case.

\section{Conclusions}

In this paper, we have presented the technical details of the derivation of the master formula \eqref{master}, which was first presented by two of us in \cite{Galda:2020epp}. In particular, we have given a detailed explanation of how to obtain this solution in the presence of the second non-local kernel arising at two loop-order in the anomalous dimension \eqref{anomalous} of the LCDA. Furthermore, we have derived an infinite set of coupled differential equations relating the logarithmic moments of the LCDA and presented their exact solution. In addition, we have worked out the solution to the evolution equations for the LCDA also in momentum space and in the diagonal (or dual) space. All results were illustrated using two model functions for the LCDA defined at the matching scale $\mu_s=1$\,GeV. In Laplace space, RG evolution to a higher scale has the effect of a global downward shift of the LCDA in the region near the origin, and it changes the location and residues of the nearest pole singularities at positive and negative values of the Laplace variable $\eta$. In momentum space, this corresponds to an increase of the value of $\lambda_B(\mu)$ as the scale $\mu$ is increased, and it has an impact on the asymptotic behavior for large and small values of the variable $\omega$. Comparing our new results obtained at NLO in RG-improved perturbation theory with the previously available LO results, we observe a genuinely small effect of the NLO contributions to the evolution equations (see Figures~\ref{fig:modelLaplace}, \ref{fig:momentumspace} and \ref{fig:Dualspace}).

In the last part of the paper, we have re-derived the explicitly scale-independent factorization formula for the convolution integral governing the $B^-\to\gamma\,\ell^-\spac\bar\nu$ decay amplitude at leading power in the heavy-quark expansion, in which all non-perturbative hadronic information is contained in the Laplace-space LCDA evaluated in the vicinity of the origin. We have evaluated this result at (approximate) NNLO in RG-improved perturbation theory, taking into account the uncertainties from variations of the matching scales $\mu_h$ and $\mu_j$. We have also discussed in detail how the results change if one adopts a different matching scale $\mu_s'>1$\,GeV. These numerical results will be of relevance to future determinations the logarithmic moments $\lambda_B$ and $\sigma_n^B$ from experimental data.

\subsection*{Acknowledgements}
We are grateful to Ben Pecjak for providing us with a MATHEMATICA implementation of the two-loop matching coefficient for the heavy-light current. This work has been supported by the Cluster of Excellence {\em Precision Physics, Fundamental Interactions, and Structure of Matter} (PRISMA$^+$\! EXC 2118/1) funded by the German Research Foundation (DFG) within the German Excellence Strategy (Project ID 39083149).

\newpage
\begin{appendix}

\section{Anomalous dimensions and RG functions}
\label{app:A}
\renewcommand{\theequation}{A.\arabic{equation}}
\setcounter{equation}{0}

We write the perturbative expansion of the various anomalous dimensions in the form
\begin{equation}
   \Gamma_{\rm cusp}(\alpha_s) = \sum_{n\ge 0}\,\Gamma_n \left( \frac{\alpha_s}{4\pi} \right)^{n+1} , \qquad
   \gamma_i(\alpha_s) = \sum_{n\ge 0}\,\gamma_{i,n} \left( \frac{\alpha_s}{4\pi} \right)^{n+1} .
\end{equation}
We evaluate the various expansion coefficients for $N_c=3$ colors. The coefficients of the cusp anomalous dimension up to four-loop order are
\begin{align}
   \Gamma_0 &= \frac{16}{3} \,, \notag\\
   \Gamma_1 &= \frac{1072}{9} - \frac{16\pi^2}{3} - \frac{160}{27}\,n_f \,, \notag\\
   \Gamma_2 &= 1960 - \frac{2144\pi^2}{9} + 352\zeta_3 + \frac{176\pi^4}{15}
    + \left( - \frac{5104}{27} + \frac{320\pi^2}{27} - \frac{832\zeta_3}{9} \right) n_f
    - \frac{64}{81}\,n_f^2 \,, \notag\\
   \Gamma_3 &= \frac{337112}{9} - \frac{178240\pi^2}{27} + 28032\zeta_3 + \frac{3608\pi^4}{5}
    - 704\pi^2\spac\zeta_3 - \frac{34496\zeta_5}{3} - 1536\zeta_3^2 - \frac{32528\pi^6}{945}  \notag\\
   &\quad + \left( -\frac{1377380}{243} + \frac{51680\pi^2}{81} - \frac{616640\zeta_3}{81}
    - \frac{2464\pi^4}{135} + \frac{1664\pi^2\spac\zeta_3}{9} + \frac{25472\zeta_5}{9} \right) n_f  \notag\\
   &\quad + \left( \frac{71500}{729} - \frac{1216\pi^2}{243} + \frac{16640\zeta_3}{81} 
    - \frac{416\pi^4}{405} \right) n_f^2 + \left( - \frac{128}{243} + \frac{256\zeta_3}{81} \right) n_f^3 \,,
\end{align}
where $n_f$ is the number of light (massless) quark flavors. The anomalous dimension $\gamma_H$ in the evolution equation for the hard function is known to three-loop order, with coefficients
\begin{equation}
\begin{aligned}
   \gamma_{H,0} &= - \frac{8}{3} \,,\\
   \gamma_{H,1} &= - \frac{2408}{27} + \frac{26\pi^2}{27} + \frac{136\zeta_3}{3}
    + \left( \frac{320}{81} + \frac{4\pi^2}{9} \right) n_f \,,\\
   \gamma_{H,2} &= -\frac{312694}{243} + \frac{8626\pi^2}{81} + \frac{53296\zeta_3}{27}
    - \frac{5726\pi^4}{1215} - \frac{6176\pi^2\spac\zeta_3}{81} - \frac{11440\zeta_5}{9} \\
   &\quad + \left( \frac{73028}{729} + \frac{260\pi^2}{27} - \frac{4832\zeta_3}{81}
    + \frac{68\pi^4}{1215} \right) n_f 
    + \left( \frac{6752}{2187} - \frac{40\pi^2}{81} - \frac{32\zeta_3}{81} \right) n_f^2 \,.
\end{aligned}
\end{equation}
Finally, the anomalous dimension $\gamma$ for the $B$-meson LCDA is known to two-loop order, with coefficients
\begin{equation}
\begin{aligned}
   \gamma_0 &= -\frac{8}{3} \,,\\
   \gamma_1 &= \frac{824}{27} - \frac{106\pi^2}{27} - \frac{200\zeta_3}{3}
    + \left( - \frac{128}{81} + \frac{20\pi^2}{27} \right) n_f \,.
\end{aligned}
\end{equation}

In the calculation of the RG functions we also need to coefficients of the QCD $\beta$-function up to four-loop order. They are given by \cite{Baikov:2016tgj}
\begin{equation}
\begin{aligned}
   \beta_0 &= 11 - \frac{2}{3}\,n_f \,, \\
   \beta_1 &= 102 - \frac{38}{3}\,n_f \,, \\
   \beta_2 &= \frac{2857}{2} - \frac{5033}{18}\,n_f + \frac{325}{54}\,n_f^2 \,, \\
   \beta_3 &= \frac{149753}{6} + 3564\spac\zeta_3 
    - \left( \frac{1078361}{162} + \frac{6508}{27}\,\zeta_{3} \right) n_f
    + \left( \frac{50065}{162} + \frac{6472}{81}\,\zeta_{3} \right) n_f^2
    + \frac{1093}{729}\,n_f^3 \,.
\end{aligned}
\end{equation}

At NLO in RG-improved perturbation theory, the RG functions $a_\Gamma$ and $S_\Gamma$ defined in \eqref{eq:aGamma} and \eqref{eq:RGFunctions} are given by
\begin{equation}
\begin{aligned}
   a_\Gamma(\nu,\mu)
   &= \frac{\Gamma_0}{2\beta_0} \left[
    \ln\frac{\alpha_s(\mu)}{\alpha_s(\nu)}
    + \left( \frac{\Gamma_1}{\Gamma_0} - \frac{\beta_1}{\beta_0} \right)
    \frac{\alpha_s(\mu) - \alpha_s(\nu)}{4\pi} + \dots \right] ,
\end{aligned}
\end{equation}
and
\begin{equation}
\begin{aligned}
   S_\Gamma(\nu,\mu) 
   &= \frac{\Gamma_0}{4\beta_0^2}\,\Bigg\{
    \frac{4\pi}{\alpha_s(\nu)} \left( 1 - \frac{1}{r} - \ln r \right)
    + \left( \frac{\Gamma_1}{\Gamma_0} - \frac{\beta_1}{\beta_0}
    \right) (1-r+\ln r) + \frac{\beta_1}{2\beta_0} \ln^2 r \\
   &\hspace{1.4cm} + \frac{\alpha_s(\nu)}{4\pi} \Bigg[ 
    \left( \frac{\beta_1\Gamma_1}{\beta_0\Gamma_0} - \frac{\beta_2}{\beta_0} 
    \right) (1-r+r\ln r)
    + \left( \frac{\beta_1^2}{\beta_0^2} - \frac{\beta_2}{\beta_0} \right)
    (1-r)\ln r \\
   &\hspace{3.2cm} - \left( \frac{\beta_1^2}{\beta_0^2} - \frac{\beta_2}{\beta_0}
    - \frac{\beta_1\Gamma_1}{\beta_0\Gamma_0} + \frac{\Gamma_2}{\Gamma_0}
    \right) \frac{(1-r)^2}{2} \Bigg] + \dots \Bigg\} \,,
\end{aligned}
\end{equation}
where $r=\alpha_s(\mu)/\alpha_s(\nu)$. The extensions of these results to NNLO can be found in equations (A.2) and (A.3) of \cite{Becher:2006mr}.

\section{Factorization in the diagonal space}
\label{app:B}
\renewcommand{\theequation}{B.\arabic{equation}}
\setcounter{equation}{0}

Using the orthonormality relation \eqref{Orthonomality}, it is straightforward to show that in diagonal space the convolution integral $I$ defined in \eqref{eq:Idef} takes the same form as in momentum space, i.e.\
\begin{equation}
   I = H(m_b,E_\gamma,\mu)\!\int_0^\infty\!\frac{d\omega}{\omega}\,
    J_{\rm diag}(-2E_\gamma\spac\omega,\mu;\rho)\,\varphi_{\rm diag}^B(\omega,\mu;\rho) \,,
\end{equation}
where the LCDAs in momentum space and in the diagonal space are related by \eqref{integraltransforms}. For the jet function, we define these transformations with the opposite transfer functions, such that
\begin{equation}
\begin{aligned}
   J_{\rm diag}(p^2,\mu;\rho) 
   &= \int_0^\infty\!\frac{dx}{\sqrt{x}}\,F_{\rm diag}^{\rm inv}(x,\mu;\rho)\,J(x p^2,\mu) \,, \\
   J(p^2,\mu) &= \int\limits_0^\infty\!\frac{dx}{\sqrt{x}}\,F_{\rm diag}(x,\mu;\rho)\,
    J_{\rm diag}\bigg(\frac{p^2}{x},\mu;\rho\bigg) \,,
\end{aligned}
\end{equation}
where the transfer functions have been given in \eqref{FPhi}. By construction, the dependence on the auxiliary scale $\rho$ cancels out in the convolution integral $I$.

The RG evolution of the jet function in the diagonal space is multiplicative, such that
\begin{equation}\label{Jdiagevol}
   J_{\rm diag}(p^2,\mu;\rho)
   = \exp\Big[ - 2 S_\Gamma(\mu_j,\mu) - a_{\gamma'}(\mu_j,\mu) \Big]
    \left( \frac{-p^2\,e^{-2\gamma_E}}{\mu_j^2} \right)^{a_\Gamma(\mu_j,\mu)}\,
    J_{\rm diag}(p^2,\mu_j;\rho) \,.
\end{equation}
The jet function at the matching scale $\mu_j$ can be written in the form \cite{Liu:2020ydl,Liu:2020eqe}
\begin{equation}\label{Jdiagmatch}
   J_{\rm diag}(p^2,\mu_j;\rho) 
   = \hat{\J}(\partial_\eta,\mu_j) \left( \frac{-p^2\spac e^{-2\gamma_E}}{\mu_j^2} \right)^\eta
    \exp\Bigg[- \int\limits_{\alpha_s(\mu_j)}^{\alpha_s(\rho)}\!\frac{d\alpha}{\beta(\alpha)}\,
    \G\big(-\eta+a_\Gamma(\mu_\alpha,\mu_j),\alpha\big) \Bigg] \Bigg|_{\eta=0} \,.
\end{equation}
The function $\hat{\J}(L_p,\mu_j)$ has been calculated at two-loop order in \cite{Liu:2020ydl}.

Combining the solutions \eqref{eq:Hevol} and \eqref{Jdiagevol}, we now obtain
\begin{equation}
\begin{aligned}
   I &= \exp\Big[ S_\Gamma(\mu_h,\mu_j) + S_\Gamma(\mu_s,\mu_j) - a_{\gamma_H}(\mu_h,\mu_j) 
    + a_\gamma(\mu_s,\mu_j) \Big]\,H(m_b,E_\gamma,\mu_h) 
    \left( \frac{2E_\gamma}{\mu_h} \right)^{-a_\Gamma(\mu_h,\mu_j)} \\
   &\quad\times \int_0^\infty\!\frac{d\omega}{\omega} 
    \left( \frac{\omega\spac e^{-2\gamma_E}}{\mu_s} \right)^{-a_\Gamma(\mu_s,\mu_j)} 
    J_{\rm diag}(-2E_\gamma\spac\omega,\mu_j;\rho)\,\varphi_{\rm diag}^B(\omega,\mu_s;\rho) \,.
\end{aligned}
\end{equation}
Notice that also this expression is manifestly independent of the factorization scale $\mu$, but contrary to \eqref{master} there is still a convolution integral remaining. As mentioned earlier, the dependence on the auxiliary scale $\rho$ cancels between the jet function and the LCDA. In Section~\ref{sec:dualSpaceSolution} we found it convenient to set $\rho=\mu_s$, so that the LCDA in the diagonal space obeys a relatively simple relation to the momentum-space LCDA, see \eqref{simplechoice}. When this is done, the jet function $J_{\rm diag}$ at the matching scale contains some large logarithms, which are resummed via \eqref{Jdiagmatch}. We find
\begin{equation}
   J_{\rm diag}(p^2,\mu_j;\mu_s) 
   = \hat{\J}(\partial_\eta,\mu_j) \left( \frac{-p^2\spac e^{-2\gamma_E}}{\mu_j^2} \right)^\eta
    \exp\Bigg[\, \int\limits_{\alpha_s(\mu_s)}^{\alpha_s(\mu_j)}\!\frac{d\alpha}{\beta(\alpha)}\,
    \G\big(-\eta+a_\Gamma(\mu_\alpha,\mu_j),\alpha\big) \Bigg] \Bigg|_{\eta=0} \,.
\end{equation}

\end{appendix}

\end{document}